\documentclass[aps,amsmath,amssymb, superscriptaddress, twocolumn, showpacs, longbibliography]{revtex4-1}
\usepackage[english]{babel}
\usepackage{graphicx}
\usepackage{amsmath}
\usepackage{amssymb}
\usepackage[outdir=./]{epstopdf}
\usepackage{amsfonts}
\usepackage{bm}
\usepackage{mathdots}
\usepackage{mathtools}
\usepackage{stmaryrd}
\usepackage{times}
\usepackage{hyperref}
\usepackage{color}
\newcommand{\ket}[1]{|#1\rangle}
\newcommand{\bra}[1]{\langle#1|}

\newcommand{\matele}[3]{\langle #1|#2|#3\rangle}
\newcommand{\I}{\mathrm{i}}

\newcommand{\ave}[1]{\langle #1 \rangle}

\newcommand{\beq}[1]{\begin{equation} #1 \end{equation}}
\newcommand{\bsplit}[1]{\begin{equation} \begin{split} #1 \end{split} \end{equation}}

\newcommand{\astcycl}{\mathrlap{\kern0.085em{\circlearrowright}}\ast}
\newcommand{\taucycl}{\mathrlap{\kern0.42em{\bullet}}\circlearrowright}
\long\def\/*#1*/{}

\newcommand{\+}{\dagger}
\def\<{\left\langle}
\def\>{\right\rangle}
\def\up{\uparrow}
\def\dn{\downarrow}

\newcommand{\vk}{\textbf{k}}
 
\newcommand{\vB}{\textbf{B}} 
\newcommand{\vG}{\textbf{G}} 
\newcommand{\vS}{\textbf{S}}
\newcommand{\vsigma}{\mbox{\boldmath$\sigma$}}
\newcommand{\vphi}{\mbox{\boldmath$\phi$}}
\newcommand{\vxi}{\mbox{\boldmath$\xi$}}
\setcitestyle{numbers,square}

\begin{document}

\title{Coupled charge and spin dynamics in a photo-excited Mott insulator}
\author{Nikolaj Bittner}
\email{nikolaj.bittner@unifr.ch}
\affiliation{Department of Physics, University of Fribourg, 1700 Fribourg, Switzerland}
\author{Denis Gole\v{z}}
\affiliation{Department of Physics, University of Fribourg, 1700 Fribourg, Switzerland}
\author{Hugo U. R. Strand}
\affiliation{Flatiron Institute, Simons Foundation, 162 Fifth Avenue. New York, NY, 10010, USA}
\author{Martin Eckstein}
\affiliation{Department of Physics, University of Erlangen-N\"urnberg, 91058 Erlangen, Germany}
\author{Philipp Werner}
\email{philipp.werner@unifr.ch}
\affiliation{Department of Physics, University of Fribourg, 1700 Fribourg, Switzerland}

\begin{abstract}
Using a nonequilibrium implementation of the extended dynamical mean field theory (EDMFT) we simulate the relaxation after photo excitation in a strongly correlated 
electron system with antiferromagnetic spin interactions. We consider the $t$-$J$ model and focus on  
the  interplay between the charge- and spin-dynamics in different excitation and doping regimes.  
The appearance of string states after a weak photo excitation manifests itself 
in a nontrivial scaling of the relaxation time with the exchange coupling 
and leads to a correlated oscillatory evolution of the kinetic energy and spin-spin correlation function. A strong 
excitation of the system, on the other hand, suppresses the spin correlations and results in a relaxation that is controlled by hole scattering. We discuss the possibility of detecting string states in 
optical and cold atom experiments. 
\end{abstract}


\maketitle
\section{Introduction}

The phase diagrams of strongly correlated materials often exhibit several competing 
phases~\cite{lee2006,keimer2015,fradkin2015}  and
a broad range of experimental probes has been used to gain insights into 
the complexity of these materials and their active degrees of freedom. 
For example, the notorious pseudogap phase  in copper based high-T$_c$ superconductors 
has been revealed and studied by nuclear magnetic resonance~\cite{warren1989,alloul1989},
optical conductivity~\cite{homes1993,timusk1999,madan2015}, and angle-resolved photo emission spectroscopy  
(ARPES)~\cite{hashimoto2014,norman1998,marshall1996}. A well documented property of underdoped cuprates 
is the tendency toward a variety of orders. In addition to superconductivity these 
include stripe and charge density wave orders~\cite{marshall1996,tranquada2004,tranquada1995},  
as well as nematic orders~\cite{hinkov2008}. Recently, a Lifshitz transition~\cite{benhabib2015} connected to the pseudo-gap phase 
has been observed in high magnetic field transport measurements under high pressure~\cite{doiron2017}. These different (incipient) orders are strongly intertwined 
and the main challenge in the field is to understand their connection to superconductivity.

The pseudo-gap phase and superconductivity in cuprates appears when holes are doped into 
a Mott insulating parent compound. Understanding the physics of doped Mott insulators is thus essential 
for the formulation of a theory of high-temperature superconductivity~\cite{lee2006}.  
A minimal model that captures the low-energy properties of cuprates is the Hubbard model. 
In the strongly interacting regime, the Fermi-Hubbard model can be mapped to 
the $t$-$J$ model~\cite{chao1977,gros1987},  
which describes the motion of holes in a spin background with anti-ferromagnetic correlations. 
The same effective theory can be obtained from the 3-band model describing the charge-transfer 
insulator set-up relevant for cuprates~\cite{emery1987}, using the insight that the doped holes 
form spin singlets~\cite{zhang1988}. Despite its apparent simplicity, the $t$-$J$ model exhibits 
a rich phase diagram with a striking similarity to that of 
cuprates~\cite{jaklivc2000finite,dagotto1994,scalapino2012}. 

Our current understanding of doped antiferromagnets is to a large extent based on numerical results. 
Exact diagonalization on small clusters~\cite{jaklivc2000finite,dagotto1994} has produced insights into 
the pairing of doped charge carriers~\cite{dagotto1994} and their interplay with short-ranged spin and 
charge fluctuations~\cite{bonca2007,tohyama2004}.  
The variational tensor network approach (iPEPS) has shown that several competing orders, namely 
$d$-wave superconductivity, charge and pair density wave states are nearly degenerate so that small changes 
in model parameters can have significant effects on the phase diagram~\cite{corboz2014}.  
Cluster extensions of dynamical mean field theory (DMFT) have been extensively used to investigate 
the pairing glue~\cite{scalapino2012,maier2006,maier2008,gull2014} and to  connect the pseudogap phase 
with the pole-like structure in the self-energy, which originates from short-range 
antiferromagnetic correlations~\cite{sakai2009,stanescu2006}. This feature in the self-energy also controls the degree of particle-hole symmetry and determines the transitions in the topology of the Fermi surface 
(Lifshitz transitions)~\cite{wu2017}.  
In the future, quantum simulators may provide additional insights into the complexity of doped antiferromagnets~\cite{mazurenko2017,hart2015,jordens2008}.  
The recent realization of N\'eel order~\cite{mazurenko2017} and 
canted antiferromagnetic states~\cite{brown2017} in cold atom experiments open the way 
to study basic questions of quantum magnetism and the effect of doping. The possibility to measure 
instantaneous high-order real-space correlation functions~\cite{hilker2017,endres2011} in these experiments provides 
an opportunity to test basic theoretical notions for doped antiferromagnets, like resonance valence bond 
(RVB) states~\cite{anderson2004},  
string states and 
Trugman paths~\cite{dagotto1994,trugman1988}, spiral states~\cite{kane1990}, or stripes, within setups 
that provide full control over the microscopic parameters~\cite{zupancic2016}. 

New insights can also be obtained by studying the nonequilibrium dynamics of charge carriers 
in these complex materials. Different intrinsic timescales allow to separate intertwined degrees of freedom 
by their temporal evolution~\cite{aoki2014_rev,allen87,giannetti2016}. 
For instance the photo-induced transition from a Mott insulator to a metal, as well as the interband relaxation and recombination of 
the charge carriers (doublons and holons) has been revealed by pump-probe optical reflectivity 
in Nd$_2$CuO$_4$ and La$_2$CuO$_4$~\cite{okamoto2011,okamoto2010}. The bosonic pairing glue has been disentangled into different contributions~\cite{dalConte2012} and it has been argued that the fast relaxation time (related to antiferromagnetic fluctuations or loop currents) is a consequence of the strong coupling between the charge and bosonic degrees of freedom responsible for pairing~\cite{dalconte2015}.  
There have been several theoretical attempts to shed light on the relaxation dynamics of photo-doped carriers in Mott insulators. The short time dynamics of holes moving in an antiferromagnetic spin background has been studied in Refs.~\onlinecite{bonca2012, golez2014, eckstein2014} while the effect of electron-phonon couplings has been investigated in Refs.~\onlinecite{golez2012a, werner2013}. A nonequilibrium extension of 
cluster DMFT~\cite{eckstein2016} and exact diagonalization calculations~\cite{zala2013,zala2014} have been used to demonstrate the ultrafast relaxation of photo-doped doublons in a system with strong antiferromagnetic short-range correlations.

Here we follow a different path by using the a nonequilibrium version of extended DMFT (EDMFT) to study the dynamics of photo-excited holes in the $t$-$J$ model. In contrast to exact diagonalization based calculations and cluster DMFT, this approach allows to study long-range spin and charge correlations, while short-range correlations may not be described as accurately. Even though the EDMFT formalism has been introduced more than a decade ago~\cite{sun2002}, most of the applications have focused on the role of non-local charge-charge interactions and the effect of dynamical screening~\cite{ayral2013,huang2014,golez2015, werner2016b}. 
Haule and co-workers~\cite{haule2002,haule2003} performed the first EDMFT simulations of the $t$-$J$ model and showed that this method captures the pseudogap phase and its connection with the 
Lifshitz transition~\cite{haule2002,haule2003}. In this work we extend the EDMFT formalism for 
the $t$-$J$ model to the nonequilibrium domain by implementing the scheme on the 
Kadanoff-Baym contour~\cite{aoki2014_rev}, and use it to study the interplay between the dynamics of spin and charge degrees of freedom. 

This paper is organized as follows: In Sec.~\ref{sec:Model} 
we introduce the two-dimensional (2D) $t$-$J$ model, which captures both  
spin and charge dynamics in the limit of an infinitely strong on-site repulsion. 
Section~\ref{sec:Method} describes the nonequilibrium implementation of extended DMFT. 
Starting from the Hubbard model we formulate the EDMFT for 
the $t$-$J$ model by implementing the projection 
to a reduced subspace without double occupation on the impurity level. In Sec.~\ref{sec:Results} 
we present simulation results for 
both the equilibrium and nonequilibrium $t$-$J$ model. In the nonequilibrium case, we focus 
on the spin and charge dynamics after weak and strong electric field excitations. 
In Sec.~\ref{sec:Summary} we summarize our results.

\section{Model}
\label{sec:Model}

We consider a strongly correlated electron system with non-local spin interactions, which is driven out of equilibrium by laser fields. 
The system is described by the single-band $t$-$J$ model~\cite{chao1977,gros1987} on a 2D square lattice
with the time-dependent Hamiltonian 
\bsplit{
  H(t)=&-\sum_{\<i,j\>\sigma}(t_h(t)\tilde c_{i\sigma}^{\dagger} \tilde c_{j\sigma}+h.c.)-\mu \sum_{i}\tilde n_i\\ 
  &+ \frac{1}{2}J \sum_{\<i,j\>} \vS_{i}\cdot\vS_{j}\ .
  \label{eq:tJ}
}
Here, the $\tilde c_{i\sigma}^{\+}$ are projected fermionic 
creation operators of an electron at site $i$  
with spin $\sigma=\{\up,\dn\}$, excluding double occupancy. They can be expressed in terms of the usual fermionic creation operators $c_{i\sigma}^{\+}$ and the density operators $n_{i\sigma}=c_{i\sigma}^\+c_{i\sigma}$ as
$\tilde c_{i\sigma}^{\+}=c_{i\sigma}^{\+} (1-n_{i\bar\sigma}),$  
and their anticommutation relation 
is given by $[\tilde c_{i\sigma},\tilde c_{j\sigma'}^{\dagger}]_+=\delta_{ij}\delta_{\sigma\sigma'}(1-n_{\bar\sigma}).$
The hopping between neighboring sites is described by $t_h(t)$, whose  
time dependence is determined by the vector potential $A(t)$ of the applied laser field. 
The projected density operator is $\tilde n_{i}=\tilde n_{i\up}+\tilde n_{i\dn}$,  with   
$\tilde n_{i\sigma}=\tilde c_{i\sigma}^\+ \tilde c_{i\sigma}$, and the hole doping is controlled by the 
chemical potential $\mu$.  Finally,   
$\vS_i=\sum_{\alpha\beta}\tilde c_{i\alpha}^\+\vsigma_{\alpha\beta}\tilde c_{i\beta}$ 
is a spin operator at site $i$ in the (Schwinger-Wigner) electron representation,  
with the vector of Pauli matrices $\vsigma_{\alpha\beta}$. The antiferromagnetic exchange parameter $J$ controls the strength of the spin interactions. 

\section{Method}
\label{sec:Method}

Dealing with projected operators within a diagrammatic formalism is in general a tedious task~\cite{ovchinnikov2004}. Here we will proceed as follows: In Sec.~\ref{edmft} we start with the extended Hubbard model with non-local spin interactions 
\bsplit{
  H(t)=&-\sum_{\<i,j\>\sigma}(t_h(t)c_{i\sigma}^{\dagger}  c_{j\sigma}+h.c.) -\mu \sum_{i} n_i \\ 
  & +U \sum_i n_{i\dn} n_{i\up} + \frac{1}{2}J \sum_{\<i,j\>} \vS_{i}\cdot\vS_{j}
  \label{eq:eHubbard}
}
and the on-site interaction $U$. 
This Hamiltonian involves the canonical fermionic operators and we can follow the usual derivation of the EDMFT approximation~\cite{ayral2013,sun2002}. The projection to the subspace without double occupancy, or equivalently $U\rightarrow \infty$, is done at the impurity level by restricting the local many-body Hilbert space, see Sec.~\ref{impurity}. Due to this projection the Dyson equation is modified, 
and we have to check if the high-energy part of the spectral weight affects the solution for the low-energy projected propagator. In Sec.~\ref{Dyson} we present a simple physical argument why this is not the case.

\subsection{Extended Dynamical Mean Field Theory}\label{edmft}
In terms of the action $S$, the grand-canonical partition function can be written as  
$\mathcal{Z}=\mathrm{Tr}[\mathcal{T_C} e^{S}]$ with $\mathcal{T_C}$ the contour-ordering operator 
on the Kadanoff-Baym contour $\mathcal{C}$~\cite{aoki2014_rev}.
For the extended Hubbard model in Eq.~\eqref{eq:eHubbard}, it can be expressed 
as a coherent-state path integral $\mathcal{Z}=\int D[c^{*}_{i}, c_i]e^{S}$ with the action 
\bsplit{
  &S[c^*,c]=-\I \int_\mathcal{C} dt dt' \Bigg\{ \sum_i U n_{i\downarrow}(t) n_{i\uparrow}(t')\delta_\mathcal{C}(t,t') \\
  &+\sum_{ij\sigma} c_{i\sigma}^{*}(t) \left[(-\I\partial_{t}
  -\mu)\delta_{ij}+t_{ij}(t)\right]\delta_\mathcal{C}(t,t') c_{j\sigma}(t')\\
   & +\frac{1}{2}  \sum_{ij} J_{ij} \vS_{i}(t)\cdot\vS_{j}(t')\delta_\mathcal{C}(t,t') \Bigg\},
  \label{eq:S}
}
where we have introduced $t_{ij}(t)=-t_h(t)\delta_{\<i,j\>}$ and $J_{ij}=J\delta_{\<i,j\>}$. 
It is convenient to decouple the spin-spin interaction part of this action by using a 
Hubbard-Stratonovich (HS) identity~\cite{negele1988} with auxiliary bosonic fields $\vphi_i$,  leading to: 

\begin{align}
\label{Eq.:Action_EDMFTwBoson}
  &S[c^*,c,\vphi]=- \I\int_\mathcal{C} \Bigg\{ \sum_{ij\sigma}c_{i\sigma}^{*}(t) [-(G_0^H)^{-1}]_{ij}(t,t') c_{j\sigma}(t')\nonumber\\
  & +\frac{1}{2}\sum_{ij} \vphi_i(t) [J^{-1}]_{ij}\delta_\mathcal{C}(t,t') \vphi_j(t') + \sum_i U n_{i\dn} n_{i\up} \nonumber\\
  & - \I \sum_{i} \vphi_{i}(t)  \delta_\mathcal{C}(t,t')\vS_{i}(t) \Bigg\}dt dt' ,
\end{align}
where the fermionic Hartree Green's function $[(G_{0}^{H})^{-1}]_{ij}=[(\I\partial_{t}+\mu)\delta_{ij}-t_{ij}]\delta_\mathcal{C}(t,t') $ has been introduced. 
The corresponding fermionic and bosonic Green's functions are
\bsplit{
  & G_{ij}(t,t')=-\I \ave{c_i(t)c_j^{\dagger}(t')}, \\
  & W_{ij}^{\alpha, \alpha'}(t,t')=\I \ave{\phi_i^\alpha(t)\phi_j^{\alpha'}(t')}, \quad \alpha, \alpha'=x,y,z, 
}
with the expectation value $\ave{\dots}=1/\mathcal{Z}\int D[c^{*}_{i}, c_i](e^{S}\dots)$. 
It should be noted that $W_{ij}$ is a tensor in spin space, which is, however, diagonal in the paramagnetic case. 
The noninteracting Green's functions (no coupling between the bosonic and fermionic fields) are given by $G_0(t,t')=G_0^H(t,t')$ and 
$W_{0,ij}(t,t')=J_{ij}\delta_\mathcal{C}(t,t')$ 
and the Dyson equations can be derived from the Baym-Kadanoff functional~\cite{ayral2013}:
\bsplit{
\label{eq:DysonLatt}
  G=G_0+G_0*\Sigma*G, \\
  W=J+J*\Pi*W,
}
where the fermionic ($\Sigma$) and bosonic ($\Pi$) self-energies were introduced and $*$ denotes the convolution on the contour $\mathcal{C}$ and a multiplication in spin space in the bosonic Dyson equation. 

We now map this lattice problem to a self-consistently determined quantum impurity problem, 
by following a nonequilibrium extended dynamical mean field theory (EDMFT) procedure 
analogous to Ref.~\onlinecite{golez2015}. Using the cavity construction~\cite{georges1996} we obtain an auxiliary impurity problem with a retarded Weiss field $\mathcal{G}_0(t,t')$ and a retarded spin interaction $\mathcal{J}(t,t'):$
\bsplit{
 S^{U}[c^*,c]_\text{eff}&=-\I \int_\mathcal{C} dt dt' \Bigg\{ \sum_{\sigma}c_{\sigma}^{*}(t) [-\mathcal{G}_0^{-1}(t,t')] c_{\sigma}(t') \\
 &+ U n_{\downarrow} n_{\uparrow} + \vS(t)[\tfrac{1}{2}\mathcal{J}(t,t')] \vS(t') \Bigg\} + \frac{1}{2}\text{Tr}[\ln\mathcal{J}].
\label{Eq.:Action_EDMFT}
}
$\mathcal{J}$ is a tensor in spin space, but by using the $SU(2)$ symmetry we can impose that the diagonal elements are identical.

\subsection{Projected impurity model}\label{impurity}

At this stage we can perform the projection to the subspace without double occupation by sending $U\rightarrow\infty$. 
The resulting impurity action reads
\bsplit{
  S=&S_0 -\I \int_\mathcal{C} dt dt' \sum_{\sigma} \tilde c^{*}_{\sigma}(t) \Delta(t,t')  \tilde  c_{\sigma}(t') \\
  &-\I \int_\mathcal{C} dt dt' \vS(t) [\tfrac12\mathcal{J}(t,t')] \vS(t').
 \label{eq:Action_Impurity}
}
 Here, we defined the local part of the action 
$S_0=-\I \int_\mathcal{C} dt dt' \left\{ \sum_{\sigma} c^{*}_{\sigma}(t) (-\I \partial_t-\mu)\delta_\mathcal{C}(t,t') 
c_{\sigma}(t')\right\}$ and 
 the hybridization function for the electrons $\Delta(t,t')$, which is related to the Weiss field by 
 $\mathcal{G}_0^{-1}(t,t')=[\I \partial_t+\mu]\delta_\mathcal{C}(t,t')-\Delta(t,t')$.

The impurity problem (\ref{eq:Action_Impurity}) can be solved using strong coupling approaches, such as the hybridization expansion~\cite{gull2011} or the non-crossing approximation (NCA) and it's extensions~\cite{grewe1981,coleman1984, eckstein2010}. The idea in the latter approaches is to introduce auxiliary pseudo-particles for  the local many body states and an additional Lagrange multiplier to fix the normalization, a detailed explanation is provided in Appendix~\ref{sec:ImpProb}. In practice we solve the impurity problem \eqref{eq:Action_Impurity} using the 
non-crossing approximation (NCA)~\cite{grewe1981, coleman1984, eckstein2010}
and obtain $G_\mathrm{imp}(t,t')$ and $W_\mathrm{imp}(t,t')$. 
Since the field $\vphi$ does not appear in the action \eqref{eq:Action_Impurity}, 
$W$ is calculated from the local spin-spin correlation function by the procedure described in Appendix~\ref{appendix_W}.
\begin{figure*}[ht]
\centering
\begin{tabular}{cp{0.1cm}c}
	\includegraphics[width=0.89\columnwidth]{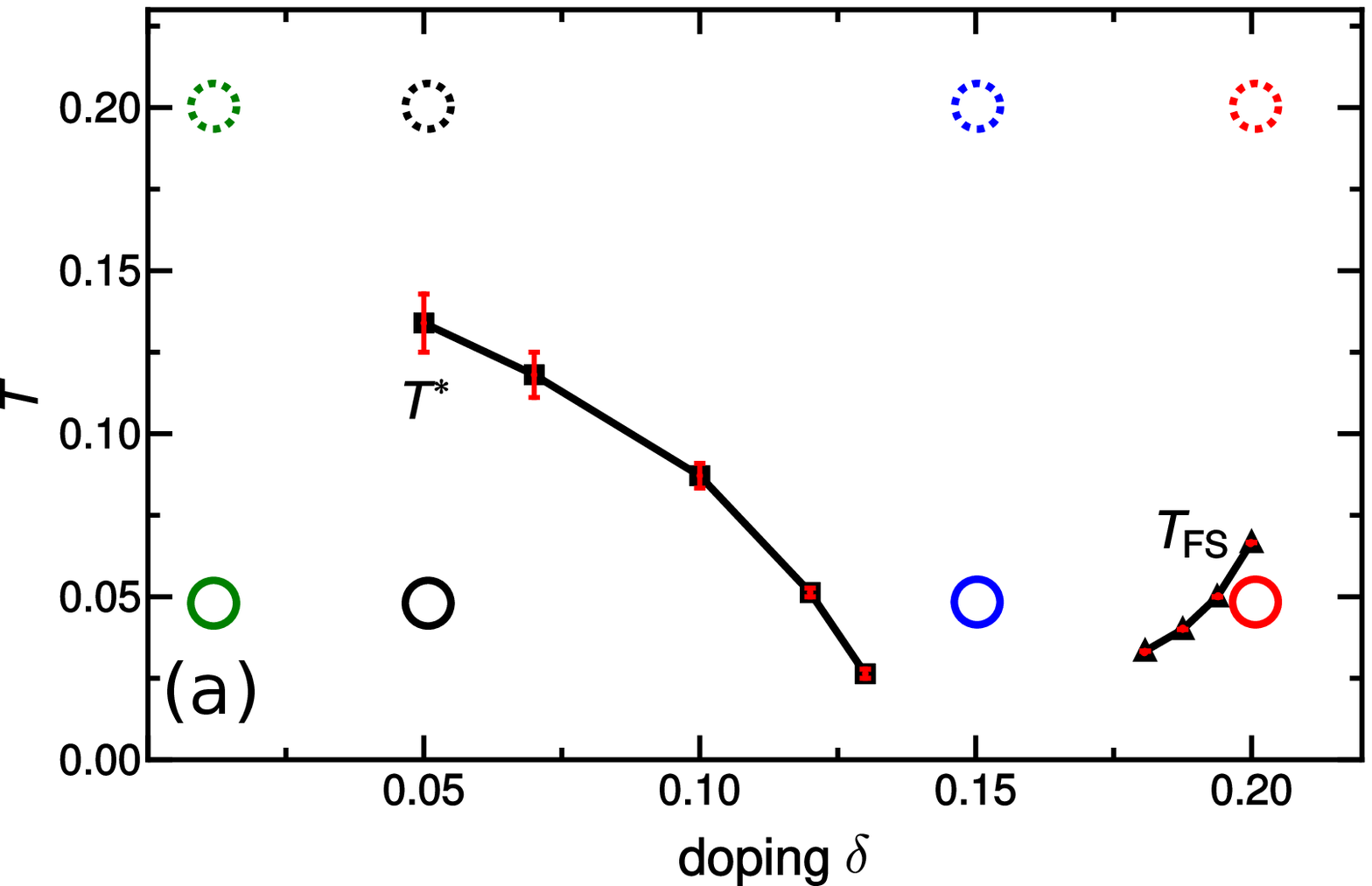}\hfill &&
	\hfill
	\includegraphics[width=1.\columnwidth]{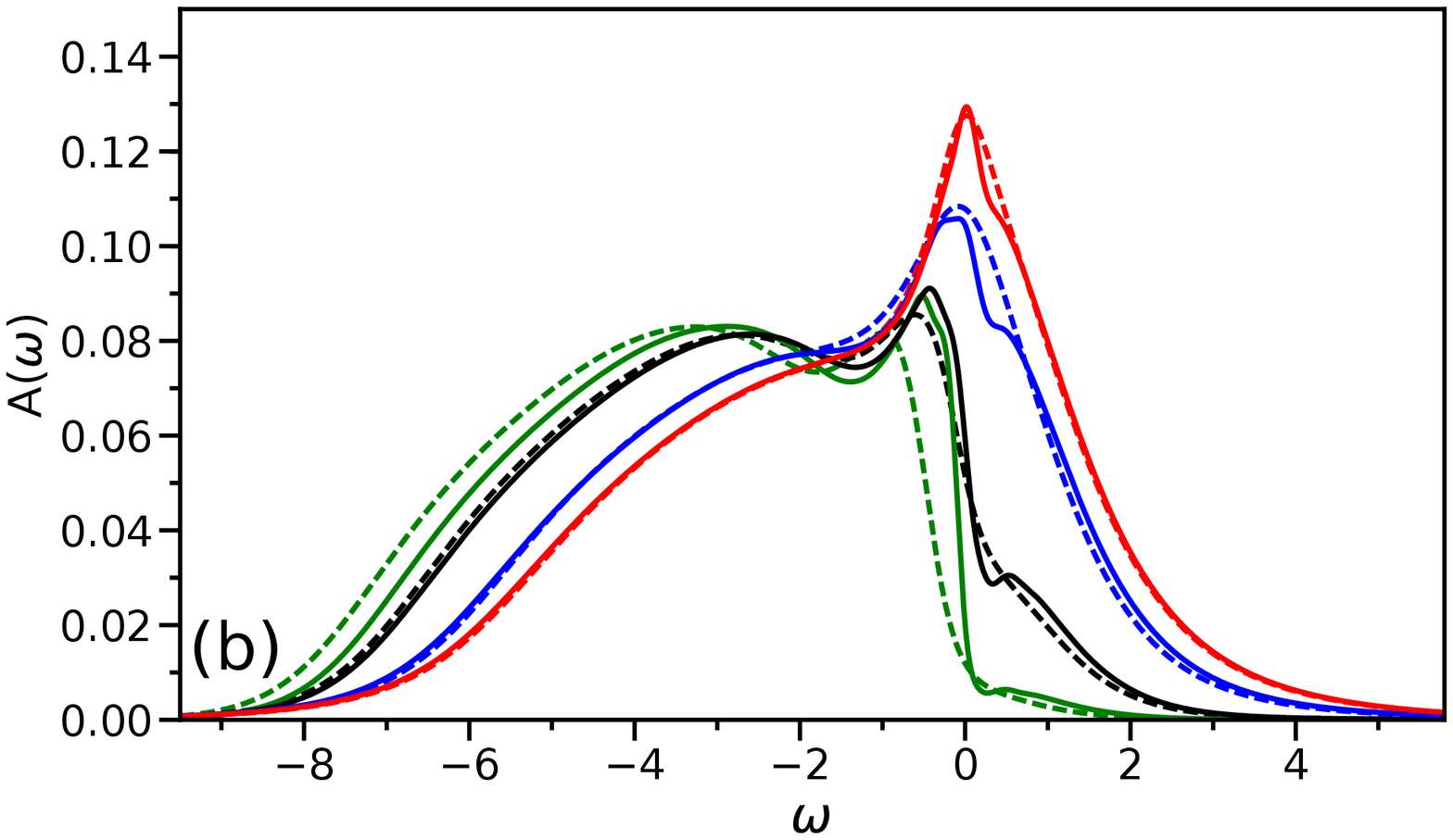}
\end{tabular}
\caption{(a) Equilibrium phase diagram of the $t$-$J$  model in the space of temperature $T$ and 
doping $\delta$. (b) Local spectral function $A(\omega)$ for $J=0.3$ and different dopings ($\delta=0.01, 0.05, 0.15, 0.20$) at temperatures $T=0.05$ (solid lines) and $T=0.2$ (dashed lines). 
The different line colors correspond to different dopings, as indicated in panel (a). 
}
\label{fig:Fig1}
\end{figure*}

\subsection{Projected Dyson equations}\label{Dyson}

Given $G_\mathrm{imp}$ and $W_\mathrm{imp}$, the fermionic self-energy $\Sigma$ and the bosonic self-energy (polarization) $\Pi$  
are obtained from the impurity Dyson equations 
\bsplit{
\label{eq:DysonImp}
	G_\mathrm{imp}&=\mathcal{G}_0+\mathcal{G}_0*\Sigma * G_\mathrm{imp},\\
	W_\mathrm{imp}&=\mathcal{J}+\mathcal{J}*\Pi * W_\mathrm{imp}.
}
These Dyson equations are valid for canonical fermionic operators, and we need to clarify how the projection performed on the impurity level modifies these expressions. At large enough $U$ we can assume that the spectral features in the self-energy $\Sigma(\omega)$ can be separated into low $\Sigma_L$ and high energy $\Sigma_H$ parts, which are well separated, i.e. $\Sigma(\omega)=\Sigma_L(\omega)+\Sigma_H(\omega).$ The fermionic Dyson equation can then be written as
\begin{align}
	&G(\omega)=\frac{1}{\omega+\mu-\epsilon_k-\Sigma_L(\omega)-\Sigma_H(\omega)}\nonumber\\
	&=\frac{1}{\omega+\mu-\epsilon_k-\Sigma_L(\omega)}+
	\frac{\Sigma_H(\omega)}{\omega+\mu-\epsilon_k-\Sigma_L(\omega)} G(\omega).
	\label{Eq.:ProDyson}
\end{align}
At low energies $\omega\ll U$, $\Sigma_H(\omega \ll U)$
is negligible, and the second term vanishes. For example, if we assume that the spectral weight at high energies can be described by a Lorentzian $\Sigma_H(\omega)=\frac{\lambda}{(\omega-U)+\I \eta}$ (the actual shape does not matter due to the energy scale separation) the second term in Eq.~(\ref{Eq.:ProDyson}) scales as $1/U$ for frequencies $\omega \ll U$ and can be neglected in the limit $U\rightarrow \infty.$
Therefore, up to $1/U$ corrections, the effective Dyson equation for the low energy degrees of freedom has the same functional form as the full Dyson equation, and we simply need to replace the full self energy $\Sigma(\omega)$ by its low energy part $\Sigma_L(\omega).$ 
In the non-equilibrium description the omission of high energy terms in the Dyson equation implies
that we are describing only the dynamics which is slower than the timescale $1/U$. 

Similar arguments hold for the lattice and impurity Dyson equations and also for the bosonic Dyson equations. The lattice self-consistency can be closed using the method discussed in Refs.~\onlinecite{aoki2014_rev,golez2015}. However, since the bosonic lattice self-consistency derived in Ref.~\onlinecite{golez2015} requires 
numerically expensive calculations, we propose here a more elegant approach, which we discuss in the next section.

\subsection{Closing the bosonic lattice self--consistency}\label{sSec:ClosureBos}

As mentioned above we extract the local bosonic self-energy $\Pi$ from 
the impurity problem. To this end, 
we compare the Dyson equations $W_\mathrm{imp}=\mathcal{J}+\mathcal{J}*\Pi*W_\mathrm{imp}$ and 
$W_\mathrm{imp}=\mathcal{J}+\mathcal{J}*\chi*\mathcal{J}$ with the spin-spin correlator $\chi(t,t')=i\<\vS(t)\ \vS(t')\>.$ After some manipulations we 
obtain the expression 
\bsplit{
\label{eq:VIEchi}
  (1+\chi*\mathcal{J})*\Pi&= \chi,
}
which is the stable version of the Volterra-Integral-Equation (VIE). Having extracted the 
self-energy $\Pi$ we can close the lattice self-consistency by solving the lattice Dyson equation
\begin{equation}
	W_\vk=J_\vk+J_\vk*\Pi*W_\vk \quad \mathrm{or} \quad (1-J_\vk*\Pi)*W_\vk=J_\vk.
\end{equation}
At this point it is useful to split $W_\vk$ into an instantaneous term $W_\vk^\delta (t)$ 
and a retarded term $W_\vk^r(t,t^\prime)$: 
$W_\vk(t,t^\prime) = W_\vk^\delta (t)\delta(t,t^\prime) + W_\vk^r(t,t^\prime)$. 
This yields the equations   
\beq{
W_\vk^{\delta}=J_\vk, \qquad [1-(J_\vk*\Pi)^r)]*W_\vk^{r}=(J_\vk*\Pi)^r*J_\vk.
}

The local bosonic Green's function $W$ is obtained from the sum over the first Brillouin zone and with this 
we can finally update the bosonic Weiss field $\mathcal{J}(t,t')$ using the impurity Dyson equation
$W=\mathcal{J}+W*\Pi*\mathcal{J}$ in the form of another stable VIE:
\beq{
  (1+W*\Pi)*\mathcal{J}=W.
}

\section{Results}
\label{sec:Results}

\subsection{Equilibrium}

First, we present equilibrium EDMFT results for the $t$-$J$ model, which were obtained using the NCA impurity solver. 
For the parameters of the system we choose $t_h=1$, and unless otherwise specified the 
exchange parameter is set to $J=0.3t_h$, which is relevant for cuprates~\cite{jaklivc2000finite}. We measure energy in units of $t_h$ and time in units 
of $\hbar/t_h$. 

\subsubsection{Phase diagram}

In Fig.~\ref{fig:Fig1}(a) we present the equilibrium phase diagram of the $t$-$J$-model 
in the space of temperature $T$ and hole concentration $\delta$. 
The equilibrium EDMFT calculations allow us to identify two transition or crossover lines, which are connected 
with (i) the onset of the pseudo-gap at $T^*(\delta)$ and (ii) the so-called Lifshitz-transition, a topological change of the Fermi-surface from hole-like to electron-like (FS) at  
$T_\mathrm{FS}(\delta)$. 
The spectral function $A(\omega)=-(1/\pi) \mathrm{Im} G^R(\omega)$ is shown in Fig.~\ref{fig:Fig1}(b) for temperatures $T=0.05$ and $T=0.2$. It represents the  
lower Hubbard band with width $\approx 8t_h$ and features a quasiparticle peak corresponding to 
holes dressed with a spin cloud.  
In the low-doping regime $\delta\lesssim 0.15$, a dip appears in the spectral function near the Fermi energy as temperature is lowered.  
The latter is a consequence of 
strong antiferromagnetic spin correlations, as discussed in more detail in connection with Fig.~\ref{fig:chiwkEq03} below, 
and thus is a manifestation of the pseudo-gap state in the EDMFT description of the $t$-$J$ model.  
We determine the pseudo-gap transition temperature by the 
appearance of this local minimum, and indicate this crossover scale in Fig.~\ref{fig:Fig1}(a) by black squares.

We next turn to the larger doping regime ($\delta>0.15$). Here, we can identify a Lifshitz transition at low temperatures, which is connected with 
a change of the FS from electron-like to hole-like. This is apparent in the spectral function $A(\omega)$ 
(see Fig.~\ref{fig:Fig1}) by a sharpening of the 
quasiparticle peak and its shift towards positive energies. 
We define the Lifshitz transition temperature $T_{FS}(\delta)$ as the temperature 
where the maximum of the quasiparticle peak of $A(\omega)$ crosses zero (i.e. shifts 
from negative to positive energies). The corresponding transition line is shown in Fig.~\ref{fig:Fig1}(a)  
by the black triangles. 
\begin{figure}[!t]
\centering
\includegraphics[width=1.0\columnwidth]{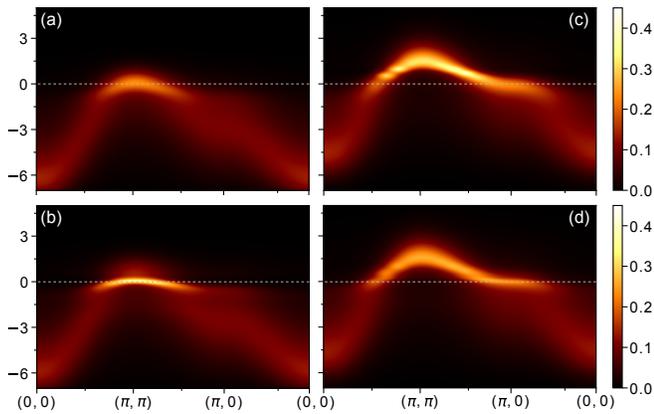}
\caption{Intensity plots of the $\vk$--dependent spectral functions $A_\vk(\omega)$ 
at $T=0.5$ (upper panels) and at $T=0.05$ (lower panels) for 
$J=0.3$ and different doping (left panels: $\delta = 0.05$, right panels: $\delta=0.20$). 
}
\label{fig:AwkEq03}
\end{figure}

\subsubsection{Spectral properties}

In Fig.~\ref{fig:AwkEq03} we plot the momentum-resolved spectral functions $A_\vk(\omega)=-(1/\pi)\mathrm{Im} G_\vk^R(\omega)$ along the diagonal and edges of the first Brillouin zone [$(0,0)\to (\pi,\pi)\to (\pi,0)\to (0,0)$]
for the underdoped ($\delta=0.05$) and overdoped ($\delta=0.20$) cases 
at low ($T=0.05<J$) and high ($T=0.5>J$) temperatures. 
In the calculations we use a grid with $16\times 16$ 
$\vk$-points and perform  an interpolation procedure.
The intensity of the spectral function is indicated by the color scale in the plots. 
Let us first focus on the 
underdoped case with $\delta=0.05$ (see Fig.~\ref{fig:AwkEq03}(a) and (b)). At low temperature 
we clearly observe a quasiparticle band with a bandwidth of $\approx 2J$, 
which is represented in the figure by 
the most intense features around the Fermi level ($\omega=0$). 
Near $\vk=(\pi,0)$ the quasiparticle band shows a flat dispersion and lies below the Fermi level. 
These observations agree with previous equilibrium studies of the $t$-$J$ 
model~\cite{tohyama2004, jaklivc2000finite}.
Furthermore, one can clearly recognize a second less coherent band with a bandwidth of $\approx 7t_h$, that resembles
the noninteracting dispersion.  
Interestingly, at $\vk\approx(\pi/2,\pi/2)$ and around $\vk=(\pi,0)$ 
there is a coexistence of both bands, i.e. there exist both renormalized quasi-particles which are strongly influenced by spin correlations and more weakly correlated incoherent states. However, increasing the temperature above $J$ 
(see Fig.~\ref{fig:AwkEq03}(a)) leads to a merging of both bands at $\vk\approx(\pi/2,\pi/2)$ and 
consequently to a so-called 
waterfall-like band dispersion similar to what has been observed in previous 
studies~\cite{Zemljic2008,Kar2011}.

Now, we turn to the overdoped case (see Fig.~\ref{fig:AwkEq03}(c) and (d)). 
Here, for $T<J$ we again observe  
sharp features corresponding to the quasiparticle band together with the second less 
coherent band. In comparison with the underdoped case, the quasiparticle band is broader and the unoccupied part of the band is weakly renormalized. Both findings qualitatively agree with 
ED calculations~\cite{jaklivc2000finite}.  
Also, at $\vk=(\pi,0)$ we find a shift of the flat quasiparticle dispersion towards the Fermi level. 
Finally, a temperature increase to $T>J$ (see Fig.~\ref{fig:AwkEq03}(c)) 
destroys the coexistence of both bands at $\vk\approx(\pi/2,\pi/2)$ and leads to a single band dispersion, 
as in the underdoped case. 
\begin{figure}[!t]
\centering
\includegraphics[width=.99\columnwidth]{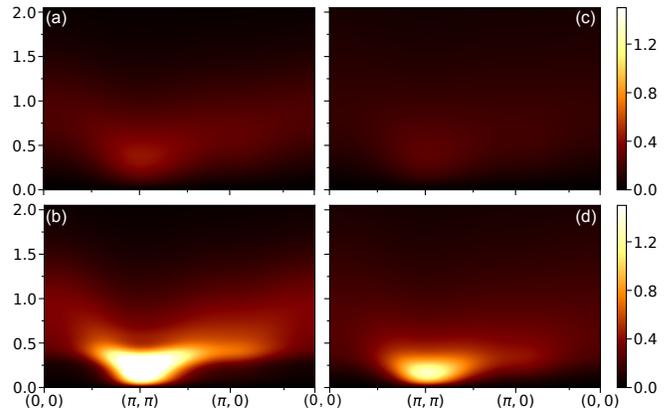}
\caption{Intensity plot of the $\vk$--dependent spin-spin correlation function Im$\chi_\vk(\omega)$ 
at $T=0.5$ (upper panels) and at $T=0.05$ (lower panels) for $J=0.3$ and different dopings (left panels: $\delta = 0.05$, right panels: $\delta=0.20$).}
\label{fig:chiwkEq03}
\end{figure}

\begin{figure*}[!t]
\centering
	\includegraphics[width=0.8\textwidth]{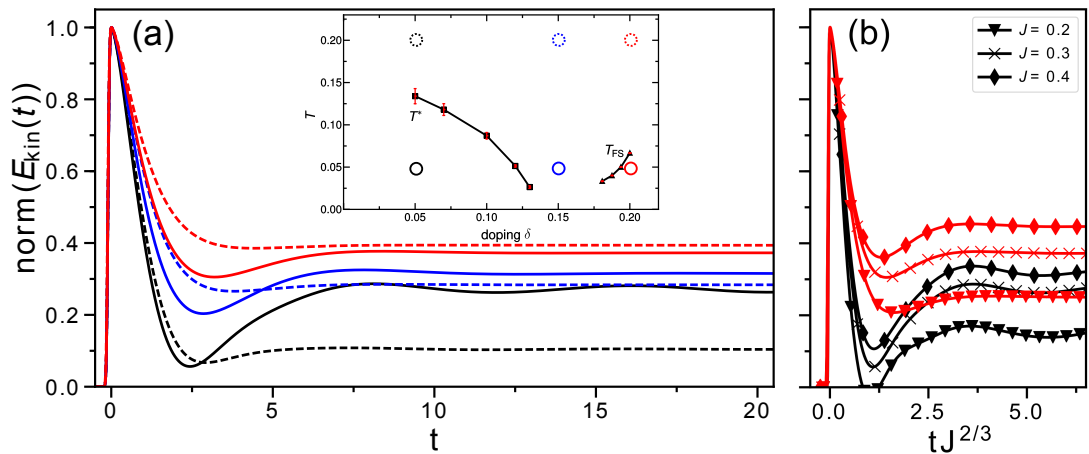}
\caption{(a) Relaxation dynamics of the normalized kinetic energy ($E_\mathrm{kin}(t)/E_\mathrm{kin}(0)$)
after a weak quench excitation 
$A(t<0)=0\rightarrow A(t\geq 0)=0.35$. 
Different lines correspond to 
different doping and temperature values as indicated in the phase diagram in the inset. 
(b) $J$-dependence of $E_\mathrm{kin}(t)$ with the rescaled time $t\to tJ^{2/3}$. Results are shown for  
$T=0.05$ and doping $\delta=0.05$ (black lines) and $\delta=0.2$ (red lines).
}
\label{fig:EkinWeak}
\end{figure*}

\subsubsection{Spin-spin correlation function}

\label{sec:SScorrfkt}

To measure the strength of the spin-spin correlations, we calculate 
the dynamical spin susceptibility
\beq{
\label{eq:chikw}
	\chi_\vk(\omega)= i\int_0^{t_\mathrm{max}} dt e^{\I \omega t}  \ave{[S^z_{\vk}(t),S^z_{-\vk}(0)]}\ ,
} 
where we take $t_\mathrm{max}=36$. 
For the evaluation of $\chi_\vk(\omega)$ 
we use a similar trick as in Sec.~\ref{sSec:ClosureBos}, and rewrite the lattice Dyson equation 
$(1+\chi_\vk * {\cal J}_\vk)*\Pi=\chi_\vk,\ $
in the form of a stable VIE for $\chi_\vk$:
\beq{
	(1-{\cal J}_\vk*\Pi)*\chi_\vk = \Pi.
}
After the solution of this equation, we perform a Fourier transformation of the resulting time-dependent $\chi_\vk(t,0)$.  
The corresponding spectra Im$\chi_\vk(\omega)$ are plotted for several dopings and temperature values 
in Fig.~\ref{fig:chiwkEq03}. 
As can be seen from the results at low temperatures, Im$\chi_{\vk}(\omega)$ exhibits low energy excitations 
near $\vk=(\pi,\pi)$ 
indicating strong antiferromagnetic correlations and a tendency to antiferromagnetic order (which is suppressed in our simulations). 
The broadening of the paramagnon is a result of fluctuations and comes from magnon-hole as well as magnon-magnon interactions. 
The strength of the spin-spin correlations decreases with increasing hole doping (compare also with the spectra in Fig.~4(f) of Ref.~\onlinecite{otsuki2013b} for the undoped case). 

\subsection{Non-Equilibrium}

Next, let us discuss the nonequilibrium dynamics of the $t$-$J$ model after an electric field quench. 
The electric field is incorporated into the Hamiltonian~\eqref{eq:tJ} by means of the Peierls substitution, i.e. 
$t_h(t)=t_h e^{iA(t)}$ with $A(t)$ the vector potential. To excite the system  
we use the fast ramp (``quench'') protocol
\beq{
	A(t)=A_0 (\mathrm{Erf}(t/\tau)+1),
}
with amplitude $A_0$ and width $\tau\approx0.07t_h$. In other words, we almost suddenly (within a small fraction of an inverse hopping time) switch the vector potential from $0$ to $A_0$ around $t=0$. 
Qualitatively similar results were also obtained for a pulse excitation (see Appendix~\ref{sSec:Pulse}).
\begin{figure}[b]
\centering
	\includegraphics[width=1.0\columnwidth]{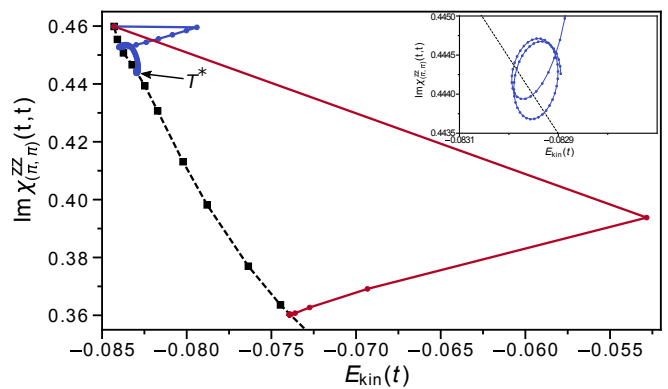}
\caption{Spin-spin correlation function at $\vk=(\pi,\pi)$ vs. kinetic energy. Black squares represent the relation between the two observables in equilibrium at different temperatures. The blue and red lines show nonequilibrium results  for $\delta=0.05$ at $T=0.05$ 
after a weak ($A(t<0)=0\rightarrow A(t\geq 0)=0.35$) and strong 
($A(t<0)=0\rightarrow A(t\geq 0)=1.4$) quench excitation, respectively. 
The inset shows a zoom of the spiral behavior after the initial relaxation for the weak excitation.
}
\label{fig:SzSzEkin}
\end{figure}
\begin{figure*}[!t]
\centering
	\includegraphics[width=1.\textwidth]{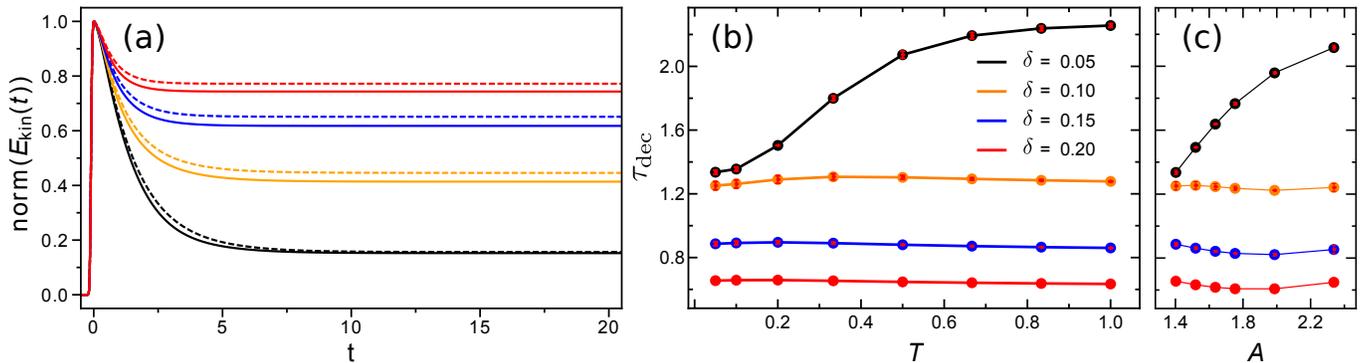}
\caption{(a) Time evolution of the normalized kinetic energy ($E_\mathrm{kin}(t)/E_\mathrm{kin}(0)$) after a strong quench excitation 
$A(t<0)=0\rightarrow A(t\geq 0)=1.4$. 
The parameters of the $t$-$J$-model used for the different lines are the same as in Fig.~\ref{fig:EkinWeak} (the additional orange lines correspond to doping $\delta=0.10$).
(b) Relaxation time $\tau_\mathrm{dec}$ vs. temperature $T$ for different doping values. 
Error bars indicate the uncertainties in the fit with Eq.~\eqref{eq:FitSE}. (c) Relaxation 
time $\tau_\mathrm{dec}$ vs. strength of the vector potential $A$ at fixed temperature $T=0.05$ and for different doping values.}
\label{fig:EkinStrong}
\end{figure*}

\subsubsection{Weak excitation}
\label{sec:WeakExc}

Figure~\ref{fig:EkinWeak}(a) illustrates the time evolution of the kinetic energy, which is normalized to its 
maximum value, after a weak quench of the vector potential $A(t<0)=0\rightarrow A(t\geq 0)=0.35$. 
Results are shown for three doping levels representing the underdoped, optimally doped and overdoped regime, and two different temperatures, 
as illustrated in the inset of the figure. 
In all cases, there is a sudden increase of the kinetic energy after the quench 
excitation and a subsequent ultrafast decrease on a timescale of a few inverse hoppings. To gain insight into the mechanism of 
this relaxation, we plot in Fig.~\ref{fig:EkinWeak}(b) the time evolution of the kinetic energy 
for different exchange parameters $J$ and dopings, using a rescaled time axis $t\to tJ^{2/3}$. 
At a fixed hole concentration the data for different $J$ show a nice collapse up to 
$tJ^{2/3}\approx 0.5$
and also a good agreement in the position of the first minimum and the subsequent oscillations. 
This observation implies that the relaxation time is larger for a system with smaller $J$ 
and hence with weaker antiferromagnetic spin correlations. 
Moreover, according to Refs.~\onlinecite{golez2014,grusdt2017} 
the $tJ^{2/3}$ scaling indicates that the reduction of the kinetic energy is associated with a local disturbance of the antiferromagnetic spin background by the creation of so-called string states~\cite{shraiman1988, dagotto1994}.

Now, let us turn back to Fig.~\ref{fig:EkinWeak}(a). 
For high temperatures we observed a simple monotonic relaxation of the kinetic energy at almost all considered hole 
concentrations, whereas at low temperatures $|E_\mathrm{kin}(t)|$ exhibits 
a minimum near $t\approx 2.5$ and a subsequent recovery with superimposed slow oscillations.

At low temperatures and in the underdoped regime, where the antiferromagnetic correlations are strong, these oscillations are particularly pronounced and long-lived. This indicates that both the recovery of the kinetic energy after the first minimum and the oscillations are the manifestation of an interplay between the charge and spin dynamics: the initially high kinetic energy of the holes is passed to the spin background (creation of string states), and the subsequent relaxation and thermalization of the locally disordered spins results in a reshuffling of kinetic and potential energy. That the spin and charge dynamics is correlated is illustrated in Fig.~\ref{fig:SzSzEkin} which plots the kinetic energy of the system against the spin-spin correlation function measured at the antiferromagnetic momentum $\bf{k}=(\pi,\pi)$, together with a line indicating the relation between these two quantities in thermal equilibrium. At low doping the time trace of the quenched system spirals around the post-quench equilibrium state (see inset), which nicely illustrates the energy flow between the electronic and spin parts of the system. 

The oscillating behavior is the direct consequence of the strong interaction between spin and charge in higher dimensional systems, in contrast to 1D chains, 
which exhibit spin-charge separation~\cite{giamarchi2004}. Based on the results of Fig.~\ref{fig:SzSzEkin} we propose that the strong coupling between spin and charge in higher dimensional systems can be unambiguously observed both in pump-probe and cold-atom experiments. An  analysis of the optical conductivity is presented in Sec.~\ref{ssSec.:Optical}. The possibility to measure the instantaneous correlation functions allows experiments with ultracold atoms to track the time dependent spin-spin and spin-charge correlation function. This ability allows the direct observation of string states, 
as also discussed Ref.~\onlinecite{grusdt2017} for a simplified $t$-$J_z$ model with Ising-like spin-spin interaction. 
In cold-atom systems the interaction is tunable and the  nontrivial $tJ^{2/3}$ scaling with time can serve as an additional indicator for the presence of the string states.

\begin{figure*}[t]
\centering
\includegraphics[width=1.0\textwidth]{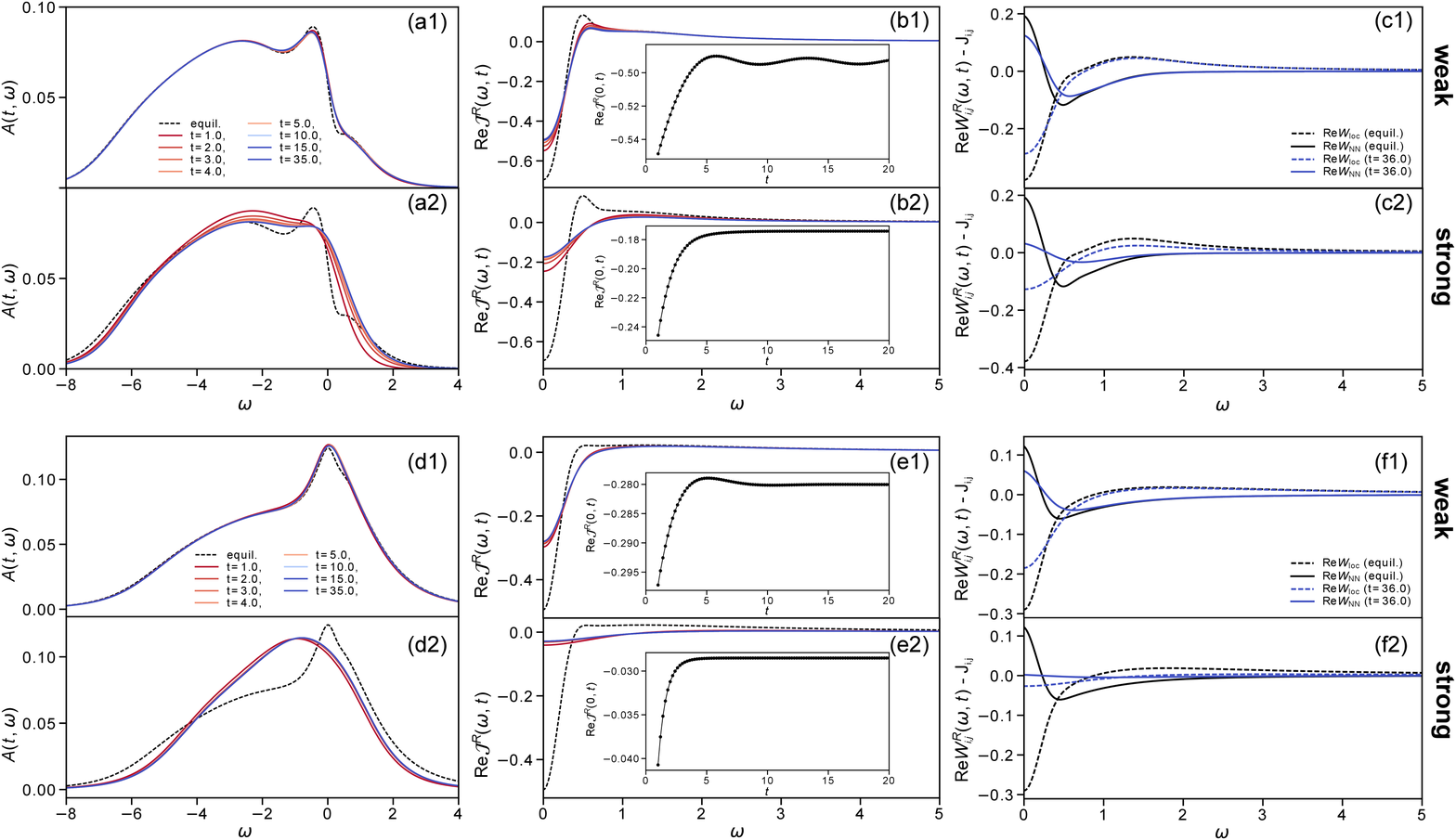}
\caption{Temporal evolution of the spectral function $A(t,\omega)$ (left panels), the real part of the impurity retarded spin-spin 
interaction Re${\cal J}^R(t,\omega)$ (middle panels) and 
the real part of the on-site (dashed lines) and nearest-neighbor (solid lines) screened effective lattice interaction Re$W^R_{ij}(t,\omega)-J_{ij}$ (right panels). 
The dynamics of the underdoped ($\delta=0.05$) spin system with $J=0.3$ at $T=0.05$ is shown after  
the weak and strong excitations in panels (a1)-(c1) and (a2)-(c2), respectively, while the temporal evolution of the overdoped 
($\delta=0.20$) system is presented in panels (d1)-(f1) and (d2)-(f2). The black dashed lines in 
the left and middle panels represent initial equilibrium results. 
The insets in the middle panels show the real part of the 
time-dependent impurity spin-spin interaction Re$\mathcal{J}^R$ in the static limit ($\omega=0$).} 
\label{fig:Wfw}
\end{figure*}

\subsubsection{Strong excitation}
\label{sec:StrExc}

Next we focus on the nonequilibrium dynamics of the $t$-$J$ model after a rather strong quench excitation 
$A(t<0)=0\rightarrow A(t\geq 0)=1.4$. 
In Fig.~\ref{fig:EkinStrong}(a) we plot again the normalized kinetic energy 
as a function of time, for the same dopings and temperatures as in Fig.~\ref{fig:EkinWeak}. 
One finds a sudden increase of the kinetic energy after the 
quench excitation and a subsequent monotonic relaxation. 
A qualitatively similar behavior of the system  
is observed if the excitation energy per hole is fixed, see Appendix~\ref{sSec:QEnHole}. 

To analyze the relaxation process 
we fit $E_\mathrm{kin}(t)$ in the time interval $t\in [0.5,43.5]$ 
using a single exponential function 
\beq{
\label{eq:FitSE}
	E_\mathrm{kin}(t)=E_\mathrm{kin}(\infty)
	+C\cdot\exp(-t/\tau_\mathrm{dec})\ ,
}
where $E_\mathrm{kin}(\infty)$
is the approximate asymptotic value of the kinetic energy, estimated at time $t=43.5$, and 
$\tau_\mathrm{dec}$ denotes the relaxation time.  
The latter is plotted as a function of temperature $T$ in Fig.~\ref{fig:EkinStrong}(b) for several doping values $\delta$. 
In the underdoped regime ($\delta<0.15$), 
the relaxation time shows a strong temperature dependence -- it decreases as temperature is lowered below $T\approx J$. 
At larger dopings, we observe that $\tau_\mathrm{dec}$ becomes almost temperature independent 
and that it decreases with increasing $\delta$. 
A similar behavior of the relaxation time of photo-excited carriers 
in an antiferromagnetically correlated background 
was also observed in Ref.~\onlinecite{eckstein2016}. This paper 
studied the two-dimensional Hubbard model in the 
large $U$ regime by means of a nonequilibrium version of cluster DMFT
and showed that the relaxation rate is proportional to the square of the nearest-neighbor spin correlations. 

This behavior of $\tau_\mathrm{dec}$ can be explained 
by the two dominant relaxation processes in our model: 
(i) relaxation through hole scattering and 
(ii) relaxation through transfer of kinetic energy to the spin background. 
Since the
short range spin correlations get weaker with increasing  temperature and increasing doping 
(see Sec.~\ref{sec:SScorrfkt}), the dominant process at high $T$ or 
large hole concentration is hole scattering. 
This implies a faster termalization of the system with increasing doping at 
fixed temperature, and hence shorter relaxation times (see Fig.~\ref{fig:EkinStrong}(b)), 
because additional holes provide additional relaxation channels. In the opposite limit of low doping 
and for temperatures roughly below $J$ the antiferromagnetic spin correlations are strong and the relaxation process (ii) dominates the dynamics. 
In this case local and collective spin excitations 
provide efficient relaxation channels that lead to a shorter relaxation time at lower temperature.  
For instance, a noticeable decrease of $\tau_\mathrm{dec}$ is found for $\delta=0.05$ as temperature is lowered  
below $T\approx J$ (see Fig.~\ref{fig:EkinStrong}(b)).

To provide additional support for the relevance of these two relaxation processes we performed calculations of the 
relaxation time for several excitation strengths. Since short-range spin correlations get weakened 
with increasing excitation strength, the relaxation through transfer of kinetic energy to the spin 
background should be suppressed. On the other hand the relaxation through hole scattering 
should get faster due to an enhanced scattering rate. 
These effects are demonstrated in Fig.~\ref{fig:EkinStrong}(c), where we fix the temperature at $T=0.05<J$ 
and vary the amplitude of the vector potential $A$ after the quench. Clearly, the relaxation time for $\delta=0.05$ 
increases with $A$, whereas it slightly decreases with increasing $A$ for larger doping.

\subsubsection{Spectral function}

In order to gain additional insights into the relaxation dynamics of the $t$-$J$ model  
we calculate the time-dependent spectral function $A(t,\omega)=-\frac{1}{\pi} \mathrm{Im}\int_{t}^{t+t_\mathrm{max}} dt' e^{\I\omega (t'-t)} G^R(t',t)$ 
from a partial Fourier transformation of the Green's function with respect to $t'$. The Fourier time window is set to $t_\mathrm{max}=22$. 
The resulting spectra at $T=0.05$ for the underdoped ($\delta=0.05$) and overdoped ($\delta=0.20$) system are shown 
in Fig.~\ref{fig:Wfw} (a1)(a2) and (d1)(d2), respectively. Panels (a1) and (d1) 
show the results after a weak excitation (as described in Sec.~\ref{sec:WeakExc}), 
whereas the strong excitation case (as described in sec~\ref{sec:StrExc}) is presented in panels (a2) and (d2). 

Let us first discuss the weak excitation regime of 
the underdoped spin system (Fig.~\ref{fig:Wfw}(a1)). In this case the quasiparticle peak gets slightly broader and its height is reduced after the excitation. The pseudo-gap (local minimum in $A(t,\omega)$) closes, but there is no significant shift of the position of the quasi-particle band and the incoherent part of the spectrum.  
On the other hand, a stronger excitation of the system 
(Fig.~\ref{fig:Wfw}(a2)) destroys the quasiparticle peak almost completely and leads to 
a substantial shift of the lower Hubbard band to higher energies. 
After the relaxation of the system at $t\gtrsim 10$ a very broad quasi-particle feature is recovered. 
The evolution of the spectral function is thus consistent with a rapid heating of the system and the thermalization at a (pulse-dependent) temperature above $T^*$.

Now, we turn to the overdoped case, which is illustrated in 
Figs.~\ref{fig:Wfw} (d1)(d2). After the weak excitation
the quasiparticle peak gets broader, whereas the 
peak position is barely changed
(see Fig.~\ref{fig:Wfw} (d1)). A further increase of the excitation 
strength (see Fig.~\ref{fig:Wfw}(d2)) 
leads to the complete melting of the quasiparticle peak 
and a simultaneous shift of the spectral weight to lower energies. The latter 
can be understood as the signature of the photo-induced Lifshitz-transition, which is associated 
with a change in the Fermi surface topology. Again, the dynamics can be understood in terms of a (pulse-dependent) heating of the system. 
(In this overdoped case, the system thermalizes already at $t\gtrsim 4$.) 

\subsubsection{Dynamics of the effective interaction}

EDMFT maps the lattice system with inter-site hopping and nonlocal antiferromagnetic spin interactions onto an effective single-site impurity problem with a hybridization function (mimicking the electron hopping processes) and an on-site retarded spin-spin interaction $\mathcal{J}$. It is interesting to look at the frequency dependence of $\mathcal{J}^R$, whose real part is plotted in the middle panels of Fig.~\ref{fig:Wfw}. The static value is negative, which indicates ferromagnetic correlations along the time axis. Robust ferromagnetic spin-spin correlations in time are the impurity model manifestation of strong antiferromagnetic correlations in space. Indeed, as we move from the underdoped (panels (b1)(b2)) to the overdoped (panels (e1)(e2)) regime, the static value in the initial equilibrium solution (dashed line) shifts closer to zero, indicating more strongly fluctuating spins and hence weaker antiferromagnetic correlations. 

The excitation of the underdoped system by a weak pulse leads to a moderate reduction in the absolute value of Re$\mathcal{J}^R(\omega=0,t)$ followed by slow oscillations (see inset) with the same frequency as previously observed in the time-evolution of the kinetic energy. This is consistent with the fact that 
antiferromagnetic correlations are still strong in an underdoped system that thermalizes at a temperature close to $T^*$ (see panel (a1)). 
After the strong excitation, the melting of the antiferromagnetic correlations is reflected in a substantially reduced  $|\text{Re}\mathcal{J}^R(\omega,t)|$ and 
an absence of oscillations in the static value. 
In the overdoped regime (panels (e1)(e2)), where the antiferromagnetic tendency is weaker already in the initial state, we do not find coherent oscillations in the evolution of $\mathcal{J}$ even after a weak excitation pulse.  

A more intuitive quantity than the retarded impurity spin-spin interaction is the screened lattice interaction $W_{ij}$. In the right hand panels of Fig.~\ref{fig:Wfw} we plot the real and imaginary parts of the on-site $W_\text{loc}$ and nearest-neighbor $W_{NN}$ in the initial state and in the thermalized state after the pulse. In the figure, we subtract the bare lattice interaction $J_{ij}$ which is equal to $J=0.3$ for the nearest neighbor component, and zero for the on-site component.  
While $W_\text{loc}$ behaves in a way analogous to the impurity interaction $\mathcal{J}$, the static value of Re$W^R_{NN}-J$ is positive, which reflects an enhanced effective antiferromagnetic nearest-neighbor coupling. The weak excitation results in a reduction of Re$W^R_{NN}-J$ by less than $50\%$, especially in the underdoped regime, while the strong excitation almost completely melts the screening contribution to the effective nearest-neighbor coupling. 

\begin{figure*}[t]
\centering
\includegraphics[width=1.0\textwidth]{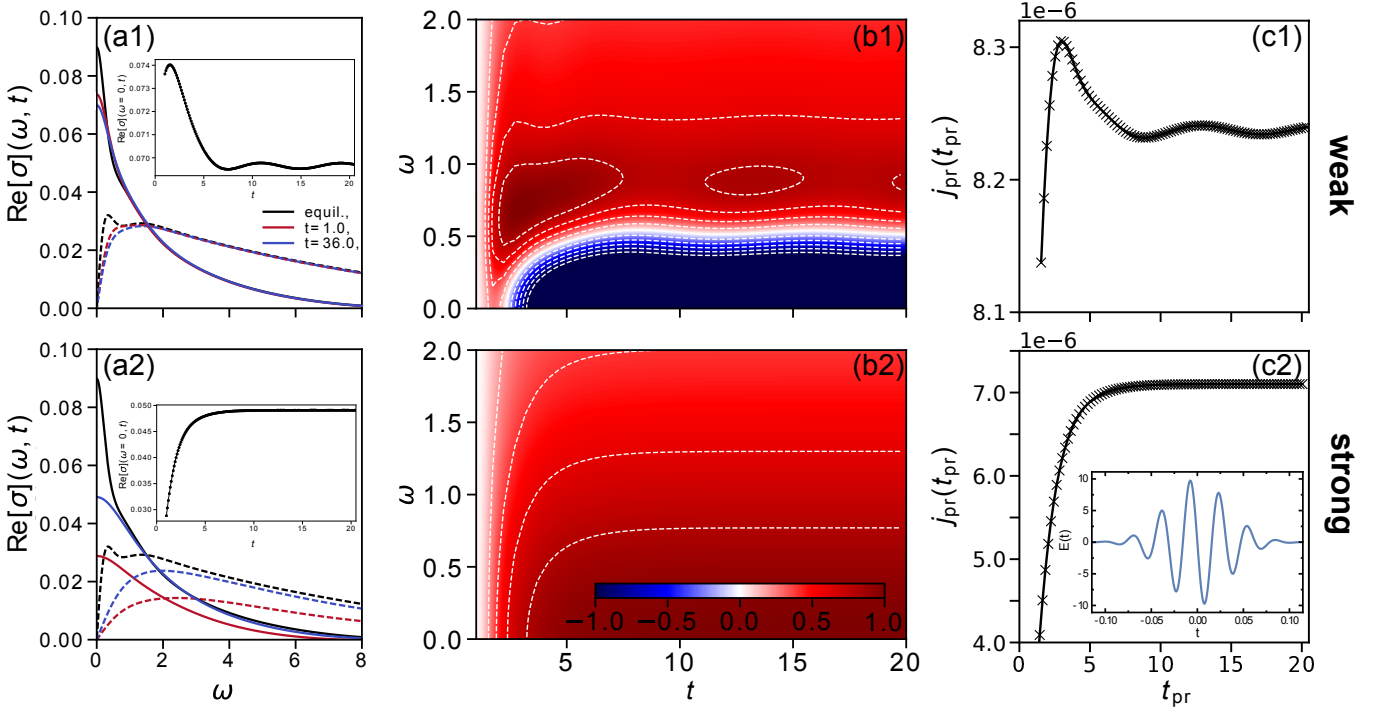}
\caption{Time-dependent optical conductivity $\sigma(\omega, t)$ 
in the underdoped case ($\delta=0.05$) at $T=0.05$ 
after (a1)-(b1) the weak quench excitation 
with $A(t<0)=0\rightarrow A(t\geq 0)=0.35$ and (a2)-(b2) the strong quench excitation 
with $A(t<0)=0\rightarrow A(t\geq 0)=1.4$. In panels (a1) and (a2) the real (solid lines) and the imaginary (dashed lines) part of the optical conductivity 
is shown for the equilibrium case (black), at $t=1$ (red), and $t=36$ (blue). 
The inset illustrates the time-dependent height of the Drude peak Re$[\sigma](\omega=0,t)$. 
In panels (b1) and (b2) the time-dependent change of the optical conductivity with respect to Re$[\sigma](\omega,t=1.0)$ is presented as an intensity plot. The intensity values are normalized to max(Re$[\sigma](\omega, t)$-Re$[\sigma](\omega,t=1.0)$). Panels (c1) and (c2) plot the current density $j_\mathrm{pr}(t_\mathrm{pr})$ induced by a probe pulse as a function of the probing delay time $t_\mathrm{pr}$ for the weak and strong excitation, respectively. The form of the probe pulse for $t_\mathrm{pr}=0$ is shown in the inset.
}
\label{fig:Sigma}
\end{figure*}

\subsubsection{Optical conductivity}
\label{ssSec.:Optical}

A fast relaxation of the Drude weight due to a strong coupling between charge and spin degrees of freedom has already been observed in optical experiments, see Ref.~\onlinecite{dalconte2015}. Here we propose that with a better time resolution additional oscillations should be revealed on top of the fast relaxation, which would serve as a ``smoking gun" for the presence of string states and strong coupling between spin and charge. The frequency of these oscillations depends on the exchange interaction, which allows to track the dependence of this microscopic parameter on external parameters such as pressure. 
The photo-induced oscillations should be strongest at weak and moderate strength of the pulse 
in order not to destroy the spin background. 

Since the clearest evidence for the appearance of 
string states was observed in the underdoped case (see Sec.~\ref{sec:WeakExc}), 
we focus in the following on the spin system with doping $\delta=0.05$.
We investigate the time evolution of the optical conductivity $\sigma(t',t)$, which for the case of a local 
self-energy reduces to a Green's function bubble~\cite{georges1996,eckstein2008,haule2003}. 
The real time dynamics is calculated using a similar procedure as described in Ref.~\onlinecite{eckstein2008}. 
From $\sigma(t',t)$ we then calculate the frequency-dependent optical conductivity by performing the partial ``forward" Fourier transformation
\beq{
	\sigma(\omega, t) = \int_t^{t+t_\mathrm{max}}dt' e^{i\omega(t'-t)}\sigma(t',t)
}
with respect to the time difference $t'-t$ at given time $t$. Here, we set $t_\mathrm{max}=22$. 
It should be noted that the nonequilibrium generalization of the f-sum rule~\cite{mahan2000}  
takes the following form~\cite{zala2014a}:
\beq{
	\int_{-\infty}^{\infty} d\omega \mathrm{Re}[\sigma](\omega,t) =\pi \sigma(t,t)=-\pi E_\mathrm{kin}(t), 
}
where $E_\mathrm{kin}(t)$ is the expectation value of the kinetic part of the Hamiltonian (\ref{eq:tJ}) measured at time $t$. 
In equilibrium the peak in the imaginary part of the optical conductivity Im$[\sigma](\omega)$ corresponds to the ``mid-infrared peak" originating from the spin fluctuations~\cite{dagotto1994,jaklivc2000finite}.

In the upper and lower panels of Fig.~\ref{fig:Sigma}, we present the time-dependent optical conductivity as 
a function of frequency for the weak (as discussed in Sec.~\ref{sec:WeakExc}) and the strong 
(as discussed in Sec.~\ref{sec:StrExc}) excitation regimes, respectively. 
In equilibrium, the real part of the optical conductivity shows a sharp Drude peak on top of a broad background. 
The Drude peak is then partially reduced after the quench in the weak excitation case (see Fig.~\ref{fig:Sigma}(a1)), 
and even more in the strong excitation limit (see Fig.~\ref{fig:Sigma}(a2)). 
Interestingly, in the latter case the Drude peak partially recovers at later times ($t=36$), i.e. after 
thermalization, whereas in the weak excitation regime it is further reduced and oscillates. The reduction of the Drude peak and subsequent oscillations are consistent with ED studies~\cite{zala2014} and we have checked that the reduced conductivity is a thermal effect. From the inset of Fig.~\ref{fig:Sigma} one can see 
that the oscillations in the $\omega=0$ value of the Drude peak  
are slightly shifted compared to the oscillations in the kinetic energy (c.f. Figs.~\ref{fig:EkinWeak} and~\ref{fig:EkinStrong}).

In Fig.~\ref{fig:Sigma}(b1) and (b2) we plot the temporal evolution of the change in the real part of the optical 
conductivity with respect to Re$[\sigma](\omega, t=1)$. The signal intensity is indicated by 
the color scale in the plots. In the weak excitation case, the height of the Drude 
peak stays suppressed after $t>2.5$ and shows an oscillating behavior. Its width gets slightly 
broader with time and also shows some oscillations. Since the spin correlations are still quite strong after the weak excitation, this time evolution can be interpreted as an energy exchange with the antiferromagnetic background.  
After a strong excitation, the weight of the Drude peak initially drops and then partially recovers after $t\approx 1$. In this case, the photo-excited system is essentially thermalized at time $t=4.5$, as confirmed by the energy distribution function. The initial decrease in the Drude weight 
may be understood as a heating effect and is consistent with simulation results for the photo-excited doped Hubbard model within single-site DMFT. The increase of the Drude weight at later times may be understood as a cooling by spin disordering, where the antiferromagnetic background plays the role of a ``heat bath".  This dynamics goes beyond the single site DMFT description of the Hubbard model, which does not capture the effect of nonlocal spin correlations, but resembles to the dynamics of a system coupled to a bosonic bath~\cite{eckstein2013}. The overall dynamics of the Drude peak is consistent with ED studies~\cite{zala2014}. 

From the optical conductivity $\sigma(t',t)$ we can calculate the current density induced by a probe pulse, a quantity that is more readily accessible in an experiment. In order to simulate realistic experimental conditions,  we describe the probe pulse by $E_\mathrm{pr}(t)=E_0\exp(-(t-t_\mathrm{pr})^2/\tau^2)\sin(\omega(t-t_\mathrm{pr}))$ and set $\omega=200$ and $\tau=0.05$. This represents a short pulse with a few cycles, as illustrated in the inset to Fig.~\ref{fig:Sigma}(c2). 
The induced current $j_\mathrm{pr}$ at time $t$ is obtained from the convolution of the optical conductivity with the probe pulse,
\beq{
	j_\mathrm{pr}(t) =\int^{t}_{0} \sigma(t,t') E_\mathrm{pr}(t')dt',
}
and the results are presented in Figs.~\ref{fig:Sigma}(c1) and (c2) for the weak and the strong excitation, respectively. Clearly, in the weak excitation case illustrated in Fig.~\ref{fig:Sigma}(c1) the induced current density $j_\mathrm{pr}$ shows  pronounced oscillations as a function of the probe pulse delay. This behavior resembles the dynamics of the kinetic energy shown in Fig.~\ref{fig:EkinWeak}. A similar agreement between the temporal evolution of the induced current and the kinetic energy is observed in the strong excitation regime (c.f. Figs.~\ref{fig:Sigma}(c2) and~\ref{fig:EkinStrong}). These observations illustrate that the nontrivial interplay between spin and charge dynamics can be directly measured in a pump-probe experiment.

\section{Summary and Conclusions}
\label{sec:Summary}

We studied the coupling between charge and spin dynamics in doped Mott insulators 
described by the two-dimensional $t$-$J$ model. To simulate the real-time evolution in 
these strongly correlated electron systems 
we used a nonequilibrium implementation of the extended 
DMFT formalism in combination with a non-crossing approximation 
impurity solver. This formalism allows to take into account non-local spin interactions, 
as well as short-ranged and long-ranged spin correlations.  

The relaxation after a weak photo-excitation exhibits 
strong correlations between the spin and charge dynamics, 
which can be related to the appearance of so-called string states. Direct evidence for this local disturbance of the 
antiferromagnetic spin background is (i) the
nontrivial scaling of the primary relaxation time with the exchange coupling $J$, and (ii) the subsequent coupled oscillations in the 
kinetic energy and spin-spin correlation function. These oscillations, which last for many periods, illustrate the flow of
energy between the spin and charge degrees of freedom.   
The latter effect is most pronounced in the underdoped regime at low temperatures ($T<J$), 
where the spin-spin correlations are the strongest, and when the excitation density is low enough such that 
the effective temperature of the underdoped system remains below or close to the pseudo-gap crossover temperature $T^*$. 
We also observed related oscillations in the height of the Drude peak 
of the optical conductivity and in the current induced by a probe pulse. This provides a path for experimentalist to 
detect string states in femto-second pump-probe studies of strongly correlated materials with strong antiferromagnetic correlations, such as cuprate superconductors.  Moreover, since the frequency of 
the oscillations depends on the exchange coupling $J$, such experiments allow 
to track this microscopic quantity as a function of macroscopic parameters. 

In the opposite limit of strong excitations we observed a rapid suppression of 
the spin-spin correlations, resulting in the relaxation of the system mainly through the hole 
scattering channel. Based on the temporal evolution of the 
spectral functions and correlation functions we interpret the dynamics of the underdoped system as a rapid heating and subsequent thermalization at $T>T^*$. Moreover, in this strong excitation regime, we observed a complete melting of the quasiparticle band after the field quench for all considered dopings. 
The closing of the peudo-gap results in a substantial shift of the lower Hubbard band to higher energies in the underdoped case, while in the overdoped case the spectral weight is shifted to lower energies. 
The latter shift results from changes in the Fermi surface topology associated with the Lifshitz transition.

On the methodological side, our study shows that the EDMFT treatment of the $t$-$J$ model can reproduce and extend previous numerical equilibrium and nonequilibrium results on doped Mott insulators. 
The formalism provides unique insights, for example into the time evolution of effective nonlocal exchange couplings, and it allowed us to reveal the conditions for strongly coupled charge and spin dynamics in two-dimensional photo-doped Mott insulators. In the future, it would be interesting to combine this nonequilibrium EDMFT approach with a cluster DMFT treatment of short-range correlations. 

\begin{acknowledgments}
	This work was supported by ERC Consolidator Grant 724103 and Swiss National Science Foundation Grant 200021-165539. The calculations have been performed on the Beo04 cluster at the University of Fribourg. 
	We thank P.~Prelov\v sek, Y.~Murakami, A.~Rosch, M.~Sch\"uler and \mbox{T.~Tohyama} for helpful discussions. 
\end{acknowledgments}

\appendix

\section{Impurity problem}
\label{sec:ImpProb}

The detailed description of the non-equilibrium impurity solver based on a combination of a 
hybridization expansion and a weak coupling expansion in powers of a retarded interaction can be found 
in Ref.~\onlinecite{golez2015}. In this appendix we explain how this technique can be adapted to 
the impurity model (\ref{eq:Action_Impurity}) which features a retarded spin-spin interaction.  

The double expansion in powers of $\Delta$ and $\mathcal{J}$ of the partition function $Z=\text{Tr}_c[T_\mathcal{C}e^\mathcal{S}]$ with action 
\bsplit{
	\mathcal{S}=&-\I\int_\mathcal{C} dt dt' \sum_\sigma c^\dagger_\sigma(t)\Delta_\sigma(t,t')c_\sigma(t')\\ 
	&+\frac{1}{2}\int_\mathcal{C}dt dt' \vS(t)\mathcal{J}(t,t')\vS(t')
+\int_\mathcal{C} dt H_\text{loc}(t)+\text{const.}
}
and $H_\text{loc}(t)=-\tilde \mu \sum_\sigma \bar n_\sigma(t) $ leads to the expression
\bsplit{
  Z  &= \sum_{n=0}^{\infty} \sum_{m=0}^{\infty} \frac{(-\I)^n}{n!} \frac{(-\I)^m}{m!}\sum_{\sigma_1\ldots\sigma_n} \text{Tr}_c\\
&\times\Bigg[\int_\mathcal{C} dt_1 \ldots  dt_{n'}  \int_\mathcal{C} d\tilde t_1 \ldots d\tilde t_{m'} T_\mathcal{C} e^{-\I\int_\mathcal{C} dt H_\text{loc}(t)} \\
&\times c^\dagger_{\sigma_1}(t_1) c_{\sigma_1}(t'_{1}) \ldots c^\dagger_{\sigma_n}(t_{n}) c_{\sigma_n}(t'_{n})\\
 &\times \left[\vS(\tilde t_{1})\cdot \vS(\tilde t'_{1})\right] \ldots \left[\vS(\tilde t_{m})\cdot\vS(\tilde t'_{m})\right]\nonumber\\
  &\times \Delta_{\sigma_1}(t_1,t_1')\ldots \Delta_{\sigma_n}(t_n,t_n')     \mathcal{J}(\tilde t_1,\tilde t_1')\ldots \mathcal{J}(\tilde t_m,\tilde t_m')
  \Bigg].
  \label{Eq.:Action_expansion}
}
 In order to evaluate the trace over the electronic configurations one can insert a complete set of states $\sum_n \ket{n}\bra{n}$ between consecutive operators $O.$ At this point we can project onto the subspace of the local many body space by restricting the sum over states and adding the Lagrange multiplier into the action to impose the normalization, namely $S=S+\lambda (\sum_n \ket{n}\bra{n}-1).$  This factors the trace into a product of impurity propagators $g$ and 
vertices for the electrons ($F^{\sigma}$) and bosons ($B$):
 \bsplit{
    g_{n}(t,t')&=-\I\matele{n}{\mathcal{T}_c e^{-\I\int_{t'}^{t} d\bar t H_\text{loc}(\bar t) }}{n}, \\
    F_{nm}^{\sigma}&=\matele{n}{c_{\sigma}}{m}, \\
    \vB_{nm}&=\matele{n}{\vS}{m},
   }
 where the spin  vertex $\vS$ mixes spin up and down states. The  main difference to the method used in Ref.~\onlinecite{golez2015} are the vertices related to the retarded spin-spin interaction. The expression for the lowest order diagram in the pseudo-particle self energy is given by
\bsplit{
  \Sigma_p(t,t') = \I[F^{\sigma} g(t,t')\bar F^{\sigma} \Delta_{\sigma}(t',t)]+\I[B g(t,t') B \mathcal{J}(t',t)].
\label{Eq.:Sigma_NCA}
}
and writing out the second term explicitly using the Pauli matrices $\sigma_{ss'}^{\alpha},$ where $\alpha=x,y,z$ and $s,s^\prime=\{\up,\dn\}$ we get
\bsplit{
  \Sigma_{p,s}^{2}(t,t') =& \I[B g(t,t') B \mathcal{J}(t',t)]\\
  =&  \I\frac{1}{4} \sum_{\alpha s'} \sigma^{\alpha
  }_{ss'} \sigma^{\alpha
  }_{s's} g_{s'}(t,t') \mathcal{J}_{\phi^{\alpha}}(t',t) \\
  =&
  \I\frac{1}{4} \sum_{\alpha s'} \sigma^{\alpha
  }_{ss'} \sigma^{\alpha
  }_{s's} g_{s'}(t,t') \mathcal{J}_{\phi}(t',t) \\=
  &\I\frac{1}{4} \sum_{s' }[2 g_{s'}(t,t')-\delta_{s,s'} g_{s}(t,t') ] \mathcal{J}_{\phi}(t',t)\\
  =&\I \frac{3}{4} g_s(t,t')\mathcal{J}_{\phi}(t',t).
\label{Eq.:Sigma_NCA_2}
}
In the step from the second to the third line we have used the spin symmetry $\mathcal{J}_{\phi^{\alpha}}=\mathcal{J}_{\phi}$, from the third to the fourth line we used the completeness relation for the Pauli matrices, namely $\vec \sigma_{ab} \vec \sigma_{cd}=2\delta_{ad}\delta_{bc}-\delta_{ab}\delta_{cd}$ and in the last line we used the fact that we are in the paramagnetic case with $g_s=g_{\bar s}.$
The explicit expressions for the NCA pseudo-particle self-energies become
\bsplit{
   \Sigma_0(t,t') = & i[(-1)G_{1\sigma}(t,t')\Lambda_{\sigma}(t',t)] ,\\
   \Sigma_{1,\sigma}(t,t') = & i[G_{0}(t,t')\Lambda_{\sigma}(t,t') +\frac{3}{4} G_{1,\sigma}(t,t')\mathcal{J}_{\phi}(t',t)  ],
  \label{Eq.:NCA}
}
where $\Sigma_0$ and $\Sigma_{1,\sigma}$ are the holon and pseudo-fermion self-energy, respectively. Surprisingly, this NCA expression for the model with retarded spin-spin interaction has the same structure as the corresponding expression in the model with retarded density-density interaction (up to a factor 3/4 which for the impurity problem can be absorbed  into a redefinition of the interaction strength). This is however a peculiarity of the NCA approximation. At the OCA level we can see the emergence of a more general structure (summation over repeated indices is assumed):
  \bsplit{
  &\Sigma_{p,s}^{4}(t,t') = \I[B g(t,t_1) B g(t_1,t_2) B g(t_2,t) \mathcal{J}(t,t_2)\mathcal{J}(t_1,t')]\\
   &= \I\frac{1}{2^4} \sigma^{\alpha}_{ss_1} \sigma^{\beta}_{s_1s_2} \sigma^{\alpha}_{s_2s_3} 
   \sigma^{\beta}_{s_3s} \\
   &\times g_{s_1}(t,t_1)g_{s_2}(t_1,t_2) g_{s_3}(t_2,t') \mathcal{J}_{\alpha}(t,t_2)\mathcal{J}_{\beta}(t_1,t')\\
  &=\I\frac{1}{2^4} [4\delta_{s,s1,s2,s3,s4}-2\delta_{s,s_3}\delta_{s_1,s_2}\\
  &\hspace{20mm}-2\delta_{s,s_1}\delta_{s_2,s_3}+\delta_{s,s1,s2,s3,s4}]\\
  &\times g_{s_1}(t,t_1) g_{s_2}(t_1,t_2) g_{s_3}(t_2,t')
  \mathcal{J}_{\alpha}(t,t_2) \mathcal{J}_{\beta}(t_1,t').
\label{Eq.:Sigma_OCA}
}
This expression cannot be mapped onto the corresponding OCA expression for the model with retarded density-density interactions. 
Note that our approach is different from the method used by Otsuki in Refs.~\onlinecite{otsuki2013a,otsuki2013b}, which employs a Lang-Firsov approach for the $S^z$-$S^z$ components of the retarded spin-spin interaction and implements a Monte Carlo sampling of the spin-flip scattering, while here we perform a weak coupling expansion in the entire retarded spin-spin interaction term. 

\begin{figure*}[!t]
\centering
\includegraphics[width=0.95\textwidth]{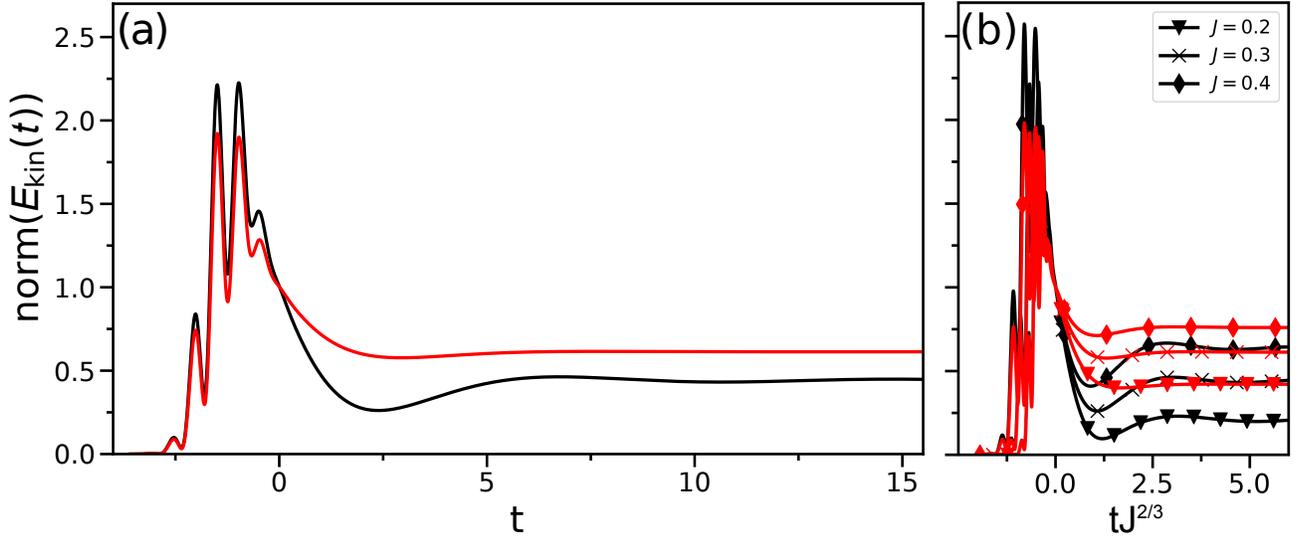}
\caption{
(a) Relaxation dynamics of the normalized kinetic energy ($E_\mathrm{kin}(t)/E_\mathrm{kin}(0)$) after a pulse 
excitation with $E_0=2.0$, $\omega=6.0$ and $\tau_w=2.1$ for $J=0.3$ in the underdoped 
($\delta=0.05$, black line) and overdoped ($\delta=0.20$, red line) cases at $T=0.05$. (b) $J$-dependence of $E_\mathrm{kin}(t)$ 
plotted as a function of the rescaled time $tJ^{2/3}$. The different line colors correspond to the same dopings as in panel (a).
} 
\label{fig:Pulse}
\end{figure*}

\subsection{Calculation of the impurity Green's function and spin susceptibility}
\label{Par.:Deriv_Gloc}

All impurity correlation functions can be expressed in terms of the pseudoparticles propagators. In order to see this we write an arbitrary 
impurity operator in the subspace 
with $Q=1$ pseudoparticles (for a more precise treatment see Ref.~\onlinecite{aoki2014_rev,eckstein2010}) as $A_i^{\dagger}=\sum_{m,n} F^i_{mn} a_{m}^{\dagger} a_n$ or $A_i=\sum_{m,n} F^i_{nm} a_{m}^{\dagger} a_n$, where $F^{i}_{mn}=\matele{m}{A_i^{\dagger}}{n}$. The impurity Green's function can then be expressed as
\begin{align}
&G_\mathrm{imp}(t,t') = - \I  \ave{c(t) c^{\dagger}(t')}= \nonumber\\
 &= -\I \sum_{n_1,n_2,m_1,m_2} \matele{n_1}{c}{m_1}  \matele{m_2}{c^{\dagger}}{n_2} \nonumber\\
 &\hspace{10mm}\times \ave{a_{n_1}^{\dagger}(t) a_{m_1}(t) a_{m_2}^{\dagger}(t') a_{n_2}(t')} \nonumber\\
 &=-\I \sum_{n_1,n_2,m_1,m_2} \matele{n_1}{c}{m_1}  \matele{m_2}{c^{\dagger}}{n_2}\nonumber\\
 &\hspace{10mm}\times[ 
 \ave{a_{n_2}(t')a_{n_1}^{\dagger}(t)}\ave{a_{m_1}(t)a_{m_2}^{\dagger}(t')}\xi_{n_1n_2} \nonumber\\
 &\hspace{10mm}+\ave{a_{m_1}(t)a_{n_1}^{\dagger}(t)}\ave{a_{n_2}(t')a_{m_2}^{\dagger}(t')}]
 \nonumber\\
 &=- \I^3  \sum_{n_1,n_2,m_1,m_2} G_{n_2,n_1}(t',t) \matele{n_1}{c}{m_1} \nonumber\\
 &\hspace{10mm}\times G_{m_1,m_2}(t,t') \matele{m_2}{c^{\dagger}}{n_2} \xi_{n_1n_2}\nonumber\\
    &= \I  \mathrm{Tr}[\vG(t',t) *  \mathbf{c} * \vG(t,t') * \mathbf{c}^\+  *\vxi],
\end{align}
where from the second to third line we have used Wick's theorem. The equal time components (loops) in
the 6th row vanish in the $Q=1$ subspace. $\vG$ is a matrix representation of the pseudoparticle propagators, 
while $\mathbf{c}$ is the matrix representation of the annihilation operators in pseudoparticle space. Furthermore, $\vxi$ is the matrix representation of 
commutator/anticommutator relations in pseudoparticle space. The same procedure can be used to evaluate the $\ave{S^z(t)S^z(t')}$ correlator (note that it is defined without a factor $-1$)  and the  final result is 
\bsplit{
\label{eq:chi}
  &\chi^{zz}=i\ave{S^z(t)S^z(t')}= \\
  &-\I \mathrm{Tr}[ \vG(t',t) * \vS^{z}*\vG(t,t')* \vS^{z} * \vxi ]= \\
  & -\frac{i}{4} \sum_s G_{s,s}(t,t') G_{s,s}(t',t)=-\frac{i}{2} G_{s}(t,t') G_{s}(t',t)  ,
}
where in the last line we only have a sum over singly occupied pseudo-particle states and assumed that spin is not important in the paramagnetic case. 
$\vS^{z}$ and $\vG$ are defined by the following matrix form:
\begin{equation}
	\vS^z=\left(
		\begin{array}{ccc}
			0&0&0\\
			0&1/2&0\\
			0&0&-1/2
		\end{array}				
		\right),\quad
		\vG=\left(
		\begin{array}{ccc}
			G_0&0&0\\
			0&G_\up&0\\
			0&0&G_\dn
		\end{array}	
		\right).
\end{equation}
A similar result can be obtained for $\chi^{xx}$ and $\chi^{yy}$. By using the matrix form of $\vS^{x}$ and $\vS^{y}$ we obtain
\begin{equation}
\vS^x=\left(
		\begin{array}{ccc}
			0&0&0\\
			0&0&1/2\\
			0&1/2&0
		\end{array}				
		\right),\quad
\vS^y=\left(
		\begin{array}{ccc}
			0&0&0\\
			0&0&-i/2\\
			0&i/2&0
		\end{array}				
		\right),
\end{equation}
and after a simple matrix multiplication one gets the following result:
\bsplit{
	\chi^{xx}=-\frac{i}{4}\left[G_\up(t,t^\prime)G_\dn(t^\prime, t)+G_\dn(t,t^\prime)G_\up(t^\prime, t)\right]=\chi^{yy}.
}
Hence, for $G_\up=G_\dn$ the spin susceptibilities are equivalent, $\chi^{xx}=\chi^{yy}=\chi^{zz}$, as it should be in the paramagnetic case.

\section{Bosonic propagator from spin-spin correlations}
\label{appendix_W}

We can calculate the bosonic propagator $W(t,t')=\I\ave{\vphi(t)\vphi(t')}$ from the spin-spin correlator $\chi(t,t')=\I\ave{\vS(t)\vS(t')} $, which can be extracted from the impurity calculation. By using the action defined in Eq.~(\ref{Eq.:Action_EDMFTwBoson}) we obtain the expression 
\bsplit{
W_\text{imp}^{ij}(t,t')&=-2\frac{\delta \ln(Z)}{\delta \mathcal{J}^{-1}_{ij}(t',t)}\\
&=2\mathcal{J}_{ik}(t,t_1) * \left[\frac{\delta \ln(Z)}{\delta \mathcal{J}(t_1,t_2)}\right]_{kl} * \mathcal{J}(t_2,t')_{lj},
}
where we have used the chain rule and the relation $\frac{\delta \mathcal{J}(t_1,t_2)}{\delta \mathcal{J}^{-1}(t',t)} = -\mathcal{J}(t_1,t') \mathcal{J}(t,t_2)$. Using Eq.~(\ref{Eq.:Action_EDMFT}) we obtain  $\frac{\delta \ln[Z]}{\delta \mathcal{J}}=-\frac{1}{2}\chi_\text{imp}+\frac{1}{2}\mathcal{J}^{-1}$ and finally arrive at  
\bsplit{
  W_\text{imp}^{ii}&=\mathcal{J}^{ii}-\mathcal{J}^{ij}\delta_{ij}*\chi_\mathrm{imp}^{jk}\delta_{ki}*\mathcal{J}^{ii} \\
&=\mathcal{J}^{ii}-\mathcal{J}^{ii}*\chi_\text{imp}^{ii}*\mathcal{J}^{ii}.
  \label{Eq.:Bosonic_from_cc}
}
Note that the spin-spin correlators $\chi_\mathrm{imp}^{ii}$, $i=x,y,z$ are equivalent for the paramagnetic case 
(see Sec.~\ref{Par.:Deriv_Gloc}). 

\section{Non-equilibrium results}

\subsection{Dynamics after a pulse excitation}
\label{sSec:Pulse}

To simulate a pulse excitation we model the electric field $E(t)=-\partial_t A(t)$ by
\beq{
	E(t)=E_0 \sin(\omega t) \exp(-4.6 t^2/\tau_w^2 )\ ,
}
with $\omega$ and $E_0$ being the frequency and field amplitude, respectively. The field has a Gaussian envelope 
of width $\tau_w$. 
For our calculations we use $\omega=6.0$ and $\tau_w=2.1$. In Fig.~\ref{fig:Pulse} we present the temporal 
evolution of the kinetic energy, which is normalized to its value at $t=0$. 
In order to perform a qualitative comparison with the results for the electric field quench
presented in Sec.~\ref{sec:WeakExc} we 
consider a weak excitation ($E_0=2.0$) of the system at $T=0.05$ in the underdoped 
($\delta=0.05$, black line) and overdoped ($\delta=0.20$, red line) regimes. From Fig.~\ref{fig:Pulse}(a) 
one can clearly see that after pumping there is a primary relaxation with 
subsequent oscillations at low doping, whereas at larger doping the oscillations are strongly suppressed. The 
primary relaxation rate and oscillations scale with $tJ^{2/3}$, as can be seen from Fig.~\ref{fig:Pulse}(b). 
All in all, these observations show a good qualitative agreement with the results for the electric field quench   
in Sec.~\ref{sec:WeakExc}.

\begin{figure}[b]
\centering
\includegraphics[width=0.95\columnwidth]{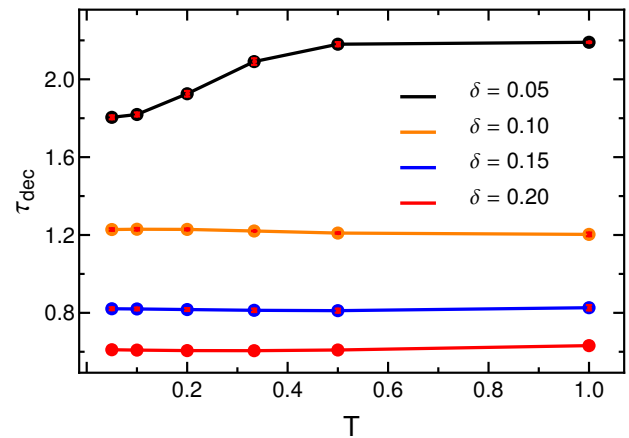}
\caption{
Relaxation time $\tau_\mathrm{dec}$ vs. temperature $T$ for different doping values 
after an electric field quench with a constant excitation energy 
per hole ($\Delta E/\delta=0.1$). 
Error bars indicate the uncertainties in the fitting with Eq.~\eqref{eq:FitSE}.
} 
\label{fig:ConstEHole}
\end{figure}

\subsection{Dynamics after an electric field quench with fixed excitation energy per hole}
\label{sSec:QEnHole}

In contrast to the case discussed in Sec.~\ref{sec:StrExc}, where we used a strong quench of 
the same intensity for all calculations, here we 
adjust the quench amplitude in order to fix the excitation energy per hole 
($\Delta E/\delta=0.1$). 
The relaxation time after excitation is again extracted using 
Eq.~\eqref{eq:FitSE}. The resulting $\tau_\mathrm{dec}$ 
is plotted in Fig.~\ref{fig:ConstEHole} as a 
function of temperature for several dopings $\delta$. 
As one can see by comparing Fig.~\ref{fig:ConstEHole} 
and Fig.~\ref{fig:EkinStrong} from Sec.~\ref{sec:StrExc}, 
the qualitative behavior 
of the relaxation rate for each doping is the same, and hence our conclusions do not 
depend on the particular excitation process. In other words, 
since in the underdoped regime ($\delta=0.05$) the spin-spin 
correlations are strong below $T<J$, one observes a lowering of $\tau_\mathrm{dec}$ by reducing temperature. In this regime, the relaxation is dominated by the interaction with the anti-ferromagnetic spin background. At the  
other considered doping values the spin-spin correlations are comparably small and the system relaxes mainly through 
hole scattering processes. 

\bibliography{Polarons,tdmft,Books}
\vspace{0cm}
\mbox{}

\end{document}